\documentclass[12pt]{article}

\usepackage{color}
\definecolor{darkblue}{rgb}{0,0,0.5}
\usepackage[unicode=true,bookmarks=false,breaklinks=false,pdfborder={0 0 1},backref=false,colorlinks=true]{hyperref}
\hypersetup{citecolor=darkblue,linkcolor=darkblue}

\title{An Approximate Kerr-Newman-like Metric Endowed with a Magnetic Dipole and Mass Quadrupole}

\author{Francisco Frutos-Alfaro}
\date{\today}

\begin{document}
\maketitle

\begin{abstract}
Approximate all-terrain spacetimes for astrophysical applications are presented. 
The metrics possess five relativistic multipole moments, namely mass, rotation, mass quadrupole, charge, and magnetic dipole moment. 
All these spacetimes approximately satisfy the Einstein-Maxwell field equations.
The first metric is generated by means of the Hoenselaers-Perj\'es method from given relativistic multipoles. 
The second metric is a perturbation of the Kerr-Newman metric, which makes it a relevant approximation for astrophysical calculations. 
The last metric is an extension of the Hartle-Thorne metric that is important for obtaining internal models of compact objects perturbatively.
The electromagnetic field is calculated using Cartan forms for locally nonrotating observers.
These spacetimes are relevant to infer properties of compact objects from astrophysical observations. 
Furthermore, the numerical implementations of these metrics are straightforward, making them versatile for simulating the potential astrophysical applications.
\end{abstract}

\section{Introduction}

\noindent
A spacetime for a real compact object is useful for many applications in astrophysics. Real compact objects possess mass, rotation, mass quadrupole and magnetic dipole, 
so that one needs a spacetime with these features. Some exact metrics containing some or all of these parameters have been obtained \cite{Bonnor,MP,AMP,Kramer,GM1,GM2,VK,GE,Bocquet}. 
Konno et al. calculated the flattening of neutron stars caused by rotation and magnetic field \cite{KTY1,KTY2}.
Using the inverse scattering method (soliton) technique of Belinskii and Zakharov \cite{BZ1,BZ2}, metrics with these characteristics have also been obtained by \cite{Chaudhuri}. 
Another technique to obtain solutions of the Einstein-Maxwell equations (EME) is the Sibgatullin method \cite{Sibgatullin} which uses the Ernst formalism \cite{Ernst1,Ernst2}; 
solutions with these features have been found by \cite{MS1,MS2,PRS,PA,MR,MMR}. 
For slowly rotating pulsars and magnetars, there are some attempts at relativistic solutions, for example \cite{MT,KK,RAM1,RAM2}. 
Another technique to generate new solutions from old ones is the Hoenselaers-Kinnersley-Xanthopoulos transformations \cite{HKX}. 
This method was used by Quevedo and Mashhoon to find a solution containing Kerr and Erez-Rosen metrics \cite{QM1,QM2}. 
The form of most of exact metrics are cumbersome to work with and are not easy to implement numerically. 
Approximations are usually sufficient to obtain relevant results, for example in \cite{OF} a ray-tracing program was used with approximate solutions compared to exact solutions.

\noindent
The detection of gravitational waves marked a historic milestone for humanity \cite{Abbott}. Phenomena such as the identification of supermassive black holes 
in the galactic center of the Milky Way, the testing of black hole spacetimes, the study of the innermost stable circular orbits (ISCO), the ring polarization and 
the observation of shadows in the Milky Way and M87 bring us closer to a better understanding of compact objects \cite{EHTC1,EHTC2,EHTC3,CFGO}. 
These findings are the motivation of this work, the metrics generated in the following sections can be applied to study these astrophysical phenomena.

\noindent
In 1989, Fodor et al. developed a procedure to obtain the relativistic multipoles using the Ernst functions for solutions of the Einstein field equations \cite{Fodor}. 
Hoenselaers and Perj\'es extended the results to electrovac solutions in 1990 \cite{Hoenselaers}. 
In addition, they proposed the reverse method, i. e. finding the metric from the relativistic multipoles. 
Pappas employed this method in 2017 to find the metric of a set of five multipoles, namely mass, spin, quadrupole, spin octupole, and mass hexadecapole \cite{Pappas}. 
To date, the electrovac algorithm has not been used to explicitly determine the metric with a set of massive, spin, and electromagnetic multipoles. 
The Hoenselaers-Perj\'es method had two mistakes, the first one was found by Sotiriou and Apostolatos \cite{SA}, and the second one by Perj\'es \cite{Perjes,FCH,CGC}. 

\noindent
In this contribution, we present approximate solutions of the EME. 
The first solution is an approximation using the improved Hoensel\-aers-Perj\'es method that employ the relativistic multipoles to generate the metric in a power series \cite{Fodor,Hoenselaers}. 
The second one is an approximation employing the Kerr-Newman metric as a seed metric, this approximation is valid up to third order in mass quadrupole and magnetic dipole 
making it attractive for computational implementations. The third one is a new version of the Hartle-Thorne metric that includes charge and the magnetic dipole. 
In this case the seed metric is Reissner-Nordstr\"om instead of Schwarzschild. 
The results were found with the help of REDUCE programs. These programs are available upon request.

\noindent
The paper is organized as follows: the second section is devoted to the Hoenselaers method to generate the metric from the relativistic multipoles. In the third section, 
a Kerr-Newman-like metric with all these parameters is developed in a perturbative way. A new version of Hartle-Thorne including charge and magnetic dipole is also generated perturbing 
a series expanded version. Some conclusions and applications are discussed in the last section. The appendices contain a summary of the relevant formulas and formalisms used in this article.

\section{Generation of the Approximate Spacetime from Relativistic Multipoles \`a la \\ Hoenselaers-Perj\'es}

\noindent
Fodor et al. developed an algorithm to compute the gravitational multipole moment of an stationary axisymmetric spacetime \cite{Fodor}.
Following this idea, Hoenselaers and Perj\'es showed in \cite{Hoenselaers} that the metric could be generated if the multipoles of the object are 
known. Sotiriou and Apostolatos corrected some typos in this Fodor's article \cite{SA}. Recently, Perj\'es found another error when computing relativistic axialsymmetric electrovacuum multipoles 
\cite{Perjes,FCH,CGC}. Moreover, Pappas found a spacetime with five relativistic multipoles, namely mass, spin, mass quadrupole, spin octupole and mass hexadecapole using this technique \cite{Pappas}. 
In this section, we generate the metric for a massive (mass, $ M $), rotating (spin, $ S $), charged object (charge, $ q_e$) endowed with a magnetic dipole ($ \mu $, magentic dipole) and 
mass quadrupole ($ M_2 $, mass quadrupole) employing this formalism. 

\noindent
The Lewis-Weyl-Papapetrou (LWP) metric in canonical cylindrical coordinates $(t, \, \rho, \, z, \, \phi)$ is given by

\begin{eqnarray}
\label{Papapetrou}
{\rm d} {s}^2 = - f ({\rm d} t - \omega {\rm d} \phi)^2
+ \frac{{\rm e}^{2 \gamma}}{f} ({\rm d} \rho^2 + {\rm d} z^2) + \frac{\rho^2}{f} {\rm d} \phi^2 ,
\end{eqnarray}

\noindent
where $f, \, \omega$ and $\gamma$ depend upon $\rho$ and $z$. This metric has to fulfill the EME with electromagnetic four-potential 
$ A_{\mu} = (- A_t, \, 0, \, 0, \, A_{\phi}) $ (see Appendix A).

\noindent
Hoenselaers and Perj\'es devised the Ernst formalism \cite{Ernst1,Ernst2} to find approximate metric components given the values of a set of relativistic multipoles. 
The Ernst formalism is based in two functions or potentials for the metric (\ref{Papapetrou}).
The first ones are the complex potentials $ {\cal E} $, and $ \Phi $. The second ones, $ \xi $ and $ q $, are defined through the latter. 
A brief overview of the Ernst formalism is in Appendix B. 

\noindent
Following this technique, the secondary Ernst functions are expanded in Taylor series

\begin{eqnarray}
\label{Papapetrou1}
{\tilde{\xi}} = \frac{1}{\bar{r}} \xi = \sum^{\infty}_{i,j = 0} a_{ij} {\bar{\rho}}^i {\bar{z}}^j , \\
{\tilde{q}} = \frac{1}{\bar{r}} q = \sum^{\infty}_{i,j = 0} b_{ij} {\bar{\rho}}^i {\bar{z}}^j , \nonumber
\end{eqnarray}

\noindent
where 

\begin{eqnarray}
\label{Papapetrou2}
{\bar{\rho}} = \frac{\rho}{\eta^2} , \quad {\bar{z}} = \frac{z}{\eta^2} , \eta = \sqrt{\rho^2 + z^2} \quad {\rm and} \quad {\bar{r}}^2 = {\bar{\rho}}^2 + {\bar{z}}^2 = r^{-2} = \eta^{-2} 
\end{eqnarray}

\noindent
with the condition that $ a_{ij} $ and $ b_{ij} $ vanish when $i$ is odd. Now, on the axis of symmetry ($ \rho = \bar{\rho} = 0 $), we have

\begin{eqnarray}
\label{Papapetrou3}
{\tilde{\xi}} (\bar{\rho} = 0) &=& \sum^{\infty}_{i= 0} m_{i} {\bar{z}}^i , \\
{\tilde{q}} (\bar{\rho} = 0) &=& \sum^{\infty}_{i= 0} q_{i} {\bar{z}}^i . \nonumber
\end{eqnarray}

\noindent
The values of $ m_{i} $ and $ q_{i} $ are related to the relativistic multipoles and to the values of the remaining non-zero values of $ a_{ij} $ and $ b_{ij} $ 
through a type of recurrence relationships (see Appendix D).
In order to obtain an approximate metric, we need to truncate these Taylor series \cite{Pappas}. In our case the truncation occurs if $ i + j \le 6 $.
Employing the recurrence relationships, the functions $ f $ and $ A_t $ are found directly ($ f = {\rm Re}[{\cal E}] + \Phi \Phi^{\star} $ and $ A_t =  {\rm Re}[{\Phi}] $). 
To find $ \omega, \, A_{\phi} $, and $\gamma $, we have to integrate (\ref{Ernstpot3}), (\ref{Ernstpot4}), (\ref{EMgrho}) or (\ref{EMgz}). 
The results for $ f, \, \omega $, and $\gamma $ are

\begin{eqnarray}
\label{Papapetrou4}
f &=& 1 - 2 M {\cal U} + (2 M^2 + q_e^2) {\cal U}^2 - 2 M^3 {\cal U}^3 + 2 M^4 {\cal U}^4 \nonumber \\
&+& (- 2 M^5 + M^3 \rho^2 + M_2 \rho^2 - 2 M_2 z^2) {\cal U}^5 \nonumber \\
&+& (- 2 M^4 \rho^2 - 2 M M_2 \rho^2 + 4 M M_2 z^2 + \mu^2 z^2 - 2 S^2 z^2) {\cal U}^6 \nonumber \\
&+& 3 M^2 (M^3 \rho^2 + M_2 \rho^2 - 2 M_2 z^2) {\cal U}^7 \\
&+& \frac{1}{28} M^2 (- 21 M^3 \rho^4 + 28 M^3 \rho^2 z^2 - 52 M_2 \rho^4 + 220 M_2 \rho^2 z^2 - 8 M_2 z^4) {\cal U}^9 , \nonumber \\
\omega &=& 2 S \rho^2 u^3 + (2 M S - \mu q_e) \rho^2 {\cal U}^4 + 4 M^2 S \rho^2 {\cal U}^5 + 2 S (2 M^3 - M_2) \rho^2 {\cal U}^6 \nonumber \\
&-& 3 M^2 S \rho^4 {\cal U}^7 + \frac{1}{2} S (- 8 M^3 + 3 M_2) \rho^4 {\cal U}^8 , \\
\gamma &=& \frac{1}{2} (- M^2 + q_e^2) \rho^2 {\cal U}^4 + (- M^4 - 3 M M_2 + 2 \mu^2 - 2 S^2) \rho^2 {\cal U}^6 \nonumber \\
&+& \frac{1}{4} (5 M^4 + 15 M M_2 - 9 \mu^2 + 9 S^2) \rho^4 {\cal U}^8 , \nonumber
\end{eqnarray}

\noindent
where $ {\cal U} = 1/\eta $.

\noindent
The components of the metric tensor are

\begin{eqnarray}
\label{Papapetrou5}
g_{tt} &=& f , \nonumber \\
g_{t \phi} &=& f \omega \nonumber \\
&=& 2 \rho^2 S {\cal U}^3 - \rho^2 (2 M S + \mu q_e) {\cal U}^4 + 4 M^2 \rho^2 S {\cal U}^5 - 2 \rho^2 S (2 M^3 + M_2) {\cal U}^6 \nonumber \\
&-& 3 M^2 \rho^4 S {\cal U}^7 + \frac{1}{2} S (8 M^3 \rho^2 + 7 M_2 \rho^2 - 8 M_2 z^2) \rho^2 {\cal U}^8 , \nonumber \\
g_{\rho \rho} &=& \frac{{\rm e}^{2 \gamma}}{f} \nonumber \\
&=& 1 + 2 M {\cal U} + (2 M^2 - q_e^2) {\cal U}^2 + 2 M^3 {\cal U}^3 + (2 M^4 - M^2 \rho^2 + q_e^2 \rho^2) {\cal U}^4 \nonumber \\
&+& (2 M^5 - 3 M^3 \rho^2 - M_2 \rho^2 + 2 M_2 z^2) {\cal U}^5 \nonumber \\
&+& (- 6 M^4 \rho^2 - 8 M M_2 \rho^2 + 4 M M_2 z^2 + 4 \mu^2 \rho^2 - \mu^2 z^2 - 4 S^2 \rho^2 + 2 S^2 z^2) {\cal U}^6 \nonumber \\
&+& 3 M^2 (- 3 M^3 \rho^2 - 5 M_2 \rho^2 + 2 M_2 z^2) {\cal U}^7 + \frac{3}{2} (2 M^4 + 5 M M_2 - 3 \mu^2 + 3 S^2) \rho^4 {\cal U}^8 \nonumber \\
&+& \frac{1}{28} M^2 (217 M^3 \rho^4 - 28 M^3 \rho^2 z^2 + 500 M_2 \rho^4 - 276 M_2 \rho^2 z^2 + 8 M_2 z^4) {\cal U}^9 , \nonumber \\
g_{z z} &=& g_{\rho \rho} , \\
g_{\phi \phi} &=& \frac{\rho^2}{f} - f \omega^2 \nonumber \\
&=& \rho^2 + 2 M \rho^2 {\cal U} + (2 M^2 - q_e^2) \rho^2 {\cal U}^2 + 2 M^3 \rho^2 {\cal U}^3 + 2 M^4 \rho^2 {\cal U}^4 \nonumber \\
&+& \rho^2 (2 M^5 - M^3 \rho^2 - M_2 \rho^2 + 2 M_2 z^2) {\cal U}^5 \nonumber \\
&+& \rho^2 (- 2 M^4 \rho^2 - 2 M M_2 \rho^2 + 4 M M_2 z^2 - \mu^2 z^2 - 4 \rho^2 S^2 + 2 S^2 z^2) {\cal U}^6 \nonumber \\
&+& 3 M^2 \rho^2 (- M^3 \rho^2 - M_2 \rho^2 + 2 M_2 z^2) {\cal U}^7 \nonumber \\
&+& \frac{1}{28} M^2 \rho^2 (21 M^3 \rho^4 - 28 M^3 \rho^2 z^2 + 52 M_2 \rho^4 - 220 M_2 \rho^2 z^2 + 8 M_2 z^4) {\cal U}^9 . \nonumber
\end{eqnarray}

\noindent
The components of the four-potential $ {A}_{\nu} = (- A_{t}, \, 0, \, 0, \, A_{\phi}) $ are

\begin{eqnarray}
\label{Papapetrou6}
A_{t} &=& q_e {\cal U} - M q_e {\cal U}^2 + \frac{1}{2} (M_2 q_e \rho^2 - 2 M_2 q_e z^2 + 2 \mu S z^2) {\cal U}^6 , \\
A_{\phi} &=& - \mu \rho^2 {\cal U}^3 + \frac{1}{2} (- M \mu + 3 q_e S) \rho^2 {\cal U}^4 + \frac{1}{2} M_2 \mu \rho^2 {\cal U}^6 - \frac{3}{8} M_2 \mu \rho^4 {\cal U}^8 . \nonumber 
\end{eqnarray}

\noindent
The electromagnetic field components for a locally non rotating observer (LNRO) can be found by using the formulas in Appendix C.

\noindent
To see which form has this metric in spherical-like coordinates, we use the Kerr-Newman mapping (Schwarzschild mapping) 

\begin{eqnarray}
\label{Papapetrou7}
\rho &=& \sqrt{\Delta} \sin{\theta} , \\
z &=& (r - M) \cos{\theta} , \nonumber 
\end{eqnarray}

\noindent
with $ a = 0 $ \cite{Frutos,Frutos2}. Then, the function $ \eta $ is

\begin{eqnarray}
\label{Papapetrou8}
\eta^2 = \Delta + (M^2 - a^2 - q_e^2) \cos^2{\theta} = r (r - 2 M) + (M^2 - q_e^2) \cos^2{\theta} .
\end{eqnarray}

\noindent
Using (\ref{Papapetrou7}), the components of the metric tensor in spherical-like coordinates $(t, \, r, \, \theta, \, \phi)$ can be obtained.
For the sake of comparison with the Kerr-Newman metric with perturbations, we change the signs of $ A_t \rightarrow - A_t $, $ S \rightarrow - S $, $ \mu \rightarrow - \mu $, 
so the metric components take the form

\begin{eqnarray}
\label{Papapetrou9}
g_{tt} &=& 1 - 2 M u + q_e^2 u^2 - 2 M_2 P_2 u^3 \nonumber \\ 
&+& \frac{1}{3} (3 \mu^2 \cos^2{\theta} - 6 M M_2 P_2 - 4 P_2 S^2 - 2 S^2) u^4 \nonumber \\
&+& \frac{2}{63} M^2 M_2 (- 35 P_2^2 - 44 P_2 + 7) u^5 , \nonumber \\
g_{t \phi} &=& (- 2 S u + \mu q_e u^2 + M_2 S (5 P_2 + 1) u^4) \sin^2{\theta} , \\
g_{rr} &=& ((r - M)^2 \sin^2{\theta} + \Delta \cos^2{\theta}) \frac{g_{\rho \rho}}{\Delta} \nonumber \\
&=& 1 + 2 M u + (4 M^2 - q_e^2) u^2 + 2 (4 M^3 + M_2 P_2) u^3 \nonumber \\
&+& \frac{1}{3} (48 M^4 + 10 M M_2 P_2^2 + 22 M M_2 P_2 - 2 M M_2 - 6 \mu^2 P_2^2 + 2 \mu^2 P_2 + \mu^2 \nonumber \\
&+& 6 S^2 P_2^2) u^4 + \frac{2}{63} M^2 (1008 M^3 + 665 M_2 P_2^2 + 548 M_2 P_2 - 133 M_2) u^5 , \nonumber \\ 
g_{\theta \theta} &=& ((r - M)^2 \sin^2{\theta} + \Delta \cos^2{\theta}) g_{z z} \nonumber \\
&=& r^2 \bigg(1 + 2 M_2 P_2 u^3 \nonumber \\
&+& \frac{1}{3} (10 M M_2 P_2^2 + 10 M M_2 P_2 - 2 M M_2 - 6 \mu^2 P_2^2 + 2 \mu^2 P_2 + \mu^2 \nonumber \\
&+& 6 S^2 P_2^2) u^4 + \frac{2}{63} M^2 M_2 (455 P_2^2 + 86 P_2 - 91) u^5 \bigg) , \nonumber \\
g_{\phi \phi} &=& r^2 \bigg(1 + 2 M_2 P_2 u^3 \nonumber \\
&+& \frac{1}{3} (18 M M_2 P_2 - 2 \mu^2 P_2 - \mu^2 + 12 P_2 S^2 - 6 S^2) u^4 \nonumber \\
&+& \frac{2}{63} M^2 M_2 (35 P_2^2 + 422 P_2 - 7) u^5 \bigg) \sin^2{\theta} , \nonumber 
\end{eqnarray}

\noindent
where $ u = 1/r $, and $ P_2 $ is the Legendre polynomial of second degree. 

\noindent
In this case, the components of the four-potential in spherical-like coordinates are

\begin{eqnarray}
\label{Papapetrou10}
A_{t} &=& - q_e u + (- \mu S \cos^2{\theta} + M_2 q_e P_2) u^4 , \\
A_{\phi} &=& \sin^2{\theta} (\mu u + \frac{3}{2} (M \mu - q_e S) u^2 - \frac{1}{4} \mu M_2 (1 + P_2) u^4) . \nonumber 
\end{eqnarray}

\noindent
The electromagnetic field components for a LNRO are (see Appendix C)

\begin{eqnarray}
\label{Papapetrou11}
E_{r} &=& q_e u^2 + 4 P_2 (- M_2 q_e + \mu S) u^5 , \nonumber \\
E_{\theta} &=& 3 \cos{\theta} \sin{\theta} (- M_2 q_e + 2 \mu S) u^5 , \\
H_{r} &=& (2 \mu u^3 + 3 (M \mu - q_e S) u^4 - 5 M_2 \mu P_2 u^6) \cos{\theta} , \nonumber \\
H_{\theta} &=& (\mu u^3 + (2 M \mu - 3 q_e S) u^4 - M_2 \mu (1 + 3 P_2) u^6) \sin{\theta} . \nonumber 
\end{eqnarray}

\noindent
The metric (\ref{Papapetrou5}) is solution of the EME up to the second order in $ M_2 $, up the third order in following parameters $ (q_{e}, \, \mu, \, S) $, and up to the sixth order in $ M $. 
For the metric functions (\ref{Papapetrou4}) not all the terms in (\ref{RM2}) and (\ref{RM3}) were kept to comply with the EME.
To keep all terms, the integration of $ \omega, \, A_\phi $ and $ \gamma $ should be improved. 
The improvement of these integrations can be done by adding new suitable terms to the metric functions $ f, \, \omega, \, \gamma $, and the four-potential functions $ A_t $, and $ A_\phi $.
The first terms in (\ref{Papapetrou11}) for $ E_{r} $, $ H_{r} $ and $ H_{\theta} $ are the usual formulas of the electric field of a charge object, and the magnetic dipole field, respectively.

\section{Generation of the Approximate Spacetime for the Kerr-Newman Metric}

\noindent
In this section, we generate a perturbed metric using the Kerr-Newman metric as a seed metric. 
To find this approximate metric, we proceed as in \cite{Frutos,Frutos2}, i.e. the magnetic dipole is included perturbatively. 
The mass quadrupole was already added in \cite{Frutos}, and in \cite{Frutos2} other parameters like mass hexadecapole, and spin octupole were taken into account.
In this case, we have to solve the EME for the Lewis metric, see Appendix A. This metric in spherical-like coordinates is 

\begin{eqnarray}
\label{Lewis}
{\rm d} s^2 = - V {\rm d} t^2 + 2 W {\rm d} t {\rm d} \phi + X {\rm d} r^2 + Y {\rm d} \theta^2 + Z {\rm d} \phi^2 ,
\end{eqnarray}

\noindent
where 

\begin{eqnarray}
\label{Lewis2}
g_{tt} &=& V = V_K + V_q + V_{\mu} , \nonumber \\
g_{t \phi} &=& W = W_K + W_q + W_{\mu} , \nonumber \\
g_{rr} &=& X = X_K + X_q + X_{\mu} , \\
g_{\theta \theta} &=& Y = Y_K + r^2 (Y_q + Y_{\mu}) , \nonumber \\
g_{\phi \phi} &=& Z = Z_K + r^2 (Z_q + Z_{\mu}) \sin^2{\theta} . \nonumber
\end{eqnarray}

\noindent
The Kerr-Newman metric potentials are given by

\begin{eqnarray}
\label{Lewis3}
V_K &=& \frac{1}{\rho^2} (\Delta - a^2 \sin^2{\theta}) , \nonumber \\
W_K &=& \frac{a}{\rho^2} (\Delta - (r^2 + a^2)) \sin^2{\theta} = - 2 (J - q_e^2) \frac{r}{\rho^2} , \nonumber \\
X_K &=& \frac{\rho^2}{\Delta} , \\
Y_K &=& \rho^2 , \nonumber \\
Z_K &=& \frac{1}{\rho^2} [(r^2 + a^2)^2 - a^2 \Delta \sin^2{\theta}] \sin^2{\theta} \nonumber
\end{eqnarray}

\noindent
with $ J = M a $,

\begin{eqnarray}
\label{Lewis4}
\rho^2 &=& r^2 + a^2 \cos^2{\theta} , \\
\Delta &=& r^2 - 2 M r + a^2 + q_e^2 . \nonumber
\end{eqnarray}

\noindent
The perturbations due to the mass quadrupole were obtained in \cite{Frutos,Frutos2}

\begin{eqnarray}
\label{Lewis5}
V_q &=& - 2 q  P_2 u^3 - 2 M q  P_2 u^4 + 2 q^2  P_2^2 u^6 , \nonumber \\
W_q &=& M a q  P^1_3 u^4 \sin{\theta} , \\
X_q &=& 2 q  P_2 u^3 + \frac{2}{7} M q (29  P_2 + 6  P_4) u^4 \nonumber \\
&+& \frac{2}{385} q^2 (44 + 55  P_2 + 36  P_4 + 250  P_6) u^6 , \nonumber \\
Y_q &=& 2 q  P_2 u^3 + \frac{6}{7} M q (5  P_2 + 2  P_4) u^4 \nonumber \\
&+& \frac{2}{385} q^2 (44 + 55  P_2 + 36  P_4 + 250  P_6) u^6 , \nonumber \\
Z_q &=& 2 q  P_2 u^3 + 6 M q  P_2 u^4 + 2 q^2  P_2^2 u^6 , \nonumber
\end{eqnarray}

\noindent
where $ u = 1/r $, $ P_n(\cos{\theta}) $, $ n = 1, \dots , 6 $, and $ P^1_3(\cos{\theta}) = (5 P_2 + 1) \sin{\theta} $ are Legendre polynomials.

\noindent
The Ansatz to include the magnetic dipole into the metric is

\begin{eqnarray}
\label{Lewis6}
V_{\mu} &=& \mu^2 (\beta_1 P_2 + \beta_2) u^4 , \nonumber \\
W_{\mu} &=& \mu q_e (\beta_3 P_2 + \beta_4) u^2 \sin^2{\theta} , \nonumber \\
X_{\mu} &=& \mu^2 (\beta_5 P_2 + \beta_6) u^4 , \\
Y_{\mu} &=& \mu^2 (\beta_7 P_2 + \beta_8) u^4 , \nonumber \\
Z_{\mu} &=& \mu^2 (\beta_9 P_2 + \beta_{10}) u^4 . \nonumber 
\end{eqnarray}

\noindent
Furthermore, the Ansatz for the electromagnetic four-potential \\ $ {A}_{\nu} = (A_{t}, \, 0, \, 0, \, A_{\phi}) $ is

\begin{eqnarray}
\label{Lewis7}
A_{t} &=& K_{t} + A_{p t} \nonumber \\
&=& K_{t} - q q_e (\alpha_1 + \alpha_2 P_1 + \alpha_3 P_2 + \alpha_4 P_3) u^4 - M a \mu (\alpha_5 P_2 + \alpha_6) u^4 , \nonumber \\
A_{\phi} &=& K_{\phi} + A_{p \phi} \nonumber \\
&=& K_{\phi}  + (\sigma_1 \mu u + \mu M (\sigma_2 P_2 + \sigma_3) u^2 + \mu q (\sigma_4 P_2 + \sigma_5) u^4) \sin^2{\theta} , 
\end{eqnarray}

\noindent
where $ {K}_{\nu} = (K_{t}, \, 0, \, 0, \, K_{\phi}) $ is the four-potential for the Kerr-Newman metric \cite{MTW}

\begin{eqnarray}
\label{Lewis8}
K_{t} &=& - \frac{q_e r}{\rho^2} , \\
K_{\phi} &=& - a K_{t} \sin^2{\theta} . \nonumber 
\end{eqnarray}

\noindent
Solving the EME (see equations from (\ref{EMLewis}) until (\ref{EMLewis2}) of Appendix A), the unkowns $ \alpha_i \, (i = 1, \dots, \, 6) $,  
$ \beta_j \, (j = 1, \dots, \, 10) $, and $ \sigma_k \, (k = 1, \dots, \, 5) $ are found solving algebraic equations:

\begin{eqnarray}
\label{Lewis9}
\beta_1 &=& \frac{2}{3} , \quad \beta_2 = \frac{1}{3} , \quad \beta_3 = 0 , \quad \beta_4 = 1 , \quad \beta_5 = 1 , \nonumber \\
\beta_6 &=& -\frac{15}{4} , \quad \beta_7 = - \frac{1}{3} , \quad \beta_8 = 1 , \quad \beta_9 = -\frac{7}{18} , \quad \beta_{10} = \frac{19}{18} , \nonumber \\
\alpha_1 &=& 0 , \quad \alpha_2 = 0 , \quad \alpha_3 = -1 , \quad \alpha_4 = \frac{2}{5} , \quad \alpha_5 = \frac{2}{3} , \quad \alpha_6 = \frac{1}{3} ,
\nonumber \\
\sigma_1 &=& 1 , \quad \sigma_2 = 0 , \quad \sigma_3 = \frac{2}{3} , \quad \sigma_4 = -\frac{1}{4} , \quad \sigma_5 = -\frac{1}{4} .
\end{eqnarray}

\noindent
The metric potentials can be written in a exponential form (except for $ W $). It is better to implement numerical codes:

\begin{eqnarray}
\label{Lewis10}
V &=& - \frac{1}{\rho^2} [(a \sin{\theta})^2 - \Delta] {\rm e}^{2 (- \psi_{q} + \psi_{1 \mu})} , \nonumber \\
W &=& \frac{a}{\rho^2}[(\Delta - (r^2 + a^2)] \sin^2{\theta} + W_{q} + W_{\mu} , \nonumber \\
X &=& \frac{\rho^2}{\Delta} {\rm e}^{2 (\chi_{q} + \chi_{1 \mu})} , \\
Y &=& \rho^2 {\rm e}^{2 (\chi_{q} + \chi_{2 \mu})} , \nonumber \\
Z &=& \frac{1}{\rho^2} [(r^2 + a^2)^2 - a^2 \Delta \sin^2{\theta}] {\rm e}^{2 (\psi_{q} + \psi_{2 \mu})} \sin^2{\theta} , \nonumber 
\end{eqnarray}

\noindent
where

\begin{eqnarray}
\label{Lewis11}
\psi_{q}     &=& q P_2 u^3 + 3 M q P_2 u^4 , \nonumber \\
\psi_{1 \mu} &=& \frac{1}{6} \mu^2 (2 P_2 + 1) u^4 = \frac{1}{2} \mu^2 u^4 \cos^2{\theta} , \nonumber \\
\psi_{2 \mu} &=& \frac{1}{36} \mu^2 (19 - 7 P_2) u^4 , \\
\chi_{q}     &=& q P_2 u^3 + \frac{1}{9} M q (15 P_2^2 + 15 P_2 - 3) u^4 \nonumber \\
&+& \frac{1}{9} q^2 (25 P_2^3 - 21 P_2^2 - 6 P_2 + 2) u^6 , \nonumber \\
\chi_{1 \mu} &=& \frac{1}{8} \mu^2 (4 P_2 - 15) u^4 , \nonumber \\
\chi_{2 \mu} &=& \frac{1}{6} \mu^2 (3 - P_2) u^4 . \nonumber
\end{eqnarray}

\noindent
The perturbed four-potential components are

\begin{eqnarray}
\label{Lewis12}
A_{p t} &=& - \frac{1}{5} q q_e (- 5 P_2 + 2 P_3) u^4 - M a \mu u^4 \cos^2{\theta} , \nonumber \\
A_{p \phi} &=& \left(\mu u + \frac{3}{2} M \mu u^2 - \frac{1}{4} \mu q (P_2 + 1) u^4 \right) \sin^2{\theta} .
\end{eqnarray}

\noindent
The electromagnetic field components for a LNRO are (see Appendix C)

\begin{eqnarray}
\label{Lewis13}
E_{r} &=& E_{K r} + E_{p r} , \nonumber \\
E_{\theta} &=& E_{K \theta} + E_{p \theta} \\
H_{r} &=& H_{K r} + H_{p r} , \nonumber \\
H_{\theta} &=& H_{K \theta} + H_{p \theta} . \nonumber
\end{eqnarray}

\noindent
The expressions for the Kerr-Newman electromagnetic fields are determined in Appendix C. The perturbation terms are found to be

\begin{eqnarray}
\label{Lewis14}
E_{p r} &=& \frac{Z_K \partial_{r} A_{p t} - W_K \partial_{r} A_{p \phi}}{\sqrt{X_K Z_K (V_K Z_K + W_K^2)}} = \partial_{r} A_{p t} + 2 J u^3 \partial_{r} A_{p \phi} \nonumber \\
&=& \frac{4}{15} q q_e u^5 (10 P_2 \cos{\theta} - 15 P_2 - 4 \cos{\theta}) + 4 a M \mu u^5 P_2 , \nonumber \\
E_{p \theta} &=& \frac{Z_K \partial_{\theta} A_{p t}- W_K \partial_{\theta} A_{p \phi}}{\sqrt{Y_K Z_K (V_K Z_K + W_K^2)}} 
= u (\partial_{\theta} A_{p t} + 2 J u^3 \partial_{\theta} A_{p \phi}) \\
&=& \frac{1}{5} q q_e u^5 \sin{\theta} (10 P_2 - 15 \cos{\theta} + 2) + 6 a M \mu u^5 \cos{\theta} \sin{\theta} , \nonumber \\
H_{p r} &=& \frac{\partial_{\theta} A_{p \phi}}{\sqrt{Y_K Z_K}} = \frac{u^2}{\sin{\theta}} \partial_{\theta} A_{p \phi} \nonumber \\
&=& 2 \mu u^3 \cos{\theta} + 3 M \mu u^4 \cos{\theta} - \mu q u^6 P_2 \cos{\theta} , \nonumber \\
H_{p \theta} &=& - \frac{\partial_{r} A_{p \phi}}{\sqrt{X_K Z_K}} = - \frac{u}{\sin{\theta}} (1 - M u) \partial_{\theta} A_{p \phi} \nonumber \\
&=& \mu u^3 \sin{\theta} + 2 M \mu u^4 \sin{\theta} - \mu q u^6 (1 + P_2) \sin{\theta} . \nonumber
\end{eqnarray}

\noindent
The spacetime components (\ref{Lewis10}) are solution of the EME up to the third order in following parameters $ (\mu, \, q) $. 

\section{Hartle-Thorne Spacetime with Charge and Magnetic Dipole}

\noindent
The Hartle-Thorne metric is an approximate solution of the Einstein equations proposed in 1968 to study the spacetime outside neutron stars, white dwarfs and supermassive stars 
with small rotation and small mass quadrupole \cite{HT}. Hartle and Thorne matched the interior and the exterior solutions, so that it is consider as an standard to know 
if a solution has physical meaning. In this section, we perturbatively include charge and magnetic dipole moments to this spacetime. 
Therefore, to include these additional features we need the Reissner-Nordstr\"om metric as seed metric. 
Other parameters like mass hexadecapole, and spin octupole were taken into account in \cite{Frutos2}.
Again, we have to solve the EME for the Lewis metric (\ref{Lewis}), see Appendix A. 
The Ansatz for the metric potentials in spherical-like coordinates is

\begin{eqnarray}
\label{HT}
V &=& \left(1 - 2 M u +  q_e^2 u^2 - \frac{2}{3} J^2 u^4 \right) {\rm e}^{2 \psi_1} , \nonumber \\
W &=& - 2 J u \sin^2{\theta} - J q P^1_3 u^4 \sin{\theta} + W_{\mu} , \\
X &=& \left(1 - 2 M u +  q_e^2 u^2 + 2 J^2 u^4 \right)^{-1} {\rm e}^{- 2 \psi_2} , \nonumber \\
Y &=& r^2 {\rm e}^{- 2 \psi_3} , \nonumber \\
Z &=& Y \sin^2{\theta} , \nonumber
\end{eqnarray}

\noindent
where $ u = 1/r $, and the functions $ \psi_i $ ($ i = 1, \, 2, \, 3 $) and $ W_{\mu} $ depend upon $ (r, \, \theta) $ and the parameters $ q $ and $ \mu $. 
The Ansatz for the functions $ \psi_i $ ($ i = 1, \, 2, \, 3 $) is

\begin{eqnarray}
\label{HT2}
\psi_1 &=& q P_2 u^3 + 3 M q P_2 u^4 - (2/3) J^2 P_2 u^4 + V_{\mu} , \\
\psi_2 &=& q P_2 u^3 + 3 M q P_2 u^4 - 8 J^2 P_2 u^4 + \frac{1}{24} q^2 (16 P_2^2 + 16 P_2 - 77) u^6 - X_{\mu} , \nonumber \\
\psi_3 &=& q P_2 u^3 + \frac{5}{2} M q P_2 u^4 - \frac{1}{2} J^2 P_2 u^4 + \frac{1}{72} q^2 (28 P_2^2 - 8 P_2 + 43) u^6 - Y_{\mu} , \nonumber
\end{eqnarray}

\noindent
where the quadrupole perturbations were obtained in \cite{Frutos2}. The functions $ V_{\mu}, \, W_{\mu}, \, X_{\mu},  \, Y_{\mu} $ have a similar form as in (\ref{Lewis6}). 
After solving the EME (see equations from (\ref{EMLewis}) until (\ref{EMLewis2}) of Appendix A), the functions $ V_{\mu}, \, W_{\mu}, \,  X_{\mu} $, and $ Y_{\mu} $ are 

\begin{eqnarray}
\label{HT3}
V_{\mu} &=& \frac{1}{2} \mu^2 u^4 \cos^2{\theta} , \nonumber \\
W_{\mu} &=&  \mu q_e u^2 \sin^2{\theta} , \\
X_{\mu} &=& \frac{1}{3} \mu^2 (P_2 - 1) u^4 , \nonumber \\
Y_{\mu} &=& - \frac{1}{6} \mu^2 P_2 u^4 . \nonumber 
\end{eqnarray}

\noindent
The components of the four-potential $ A_{\nu} $ are

\begin{eqnarray}
\label{HT4}
A_{t} &=& - q_e u - J \mu u^4 \cos^2{\theta} - q q_e P_2 u^4 , \\
A_{\phi} &=& \left(\mu u - \frac{3}{2} J q_e u^2 + \frac{3}{2} M \mu u^2 + \frac{1}{4} \mu q (P_2 + 1) u^4 \right) \sin^2{\theta} . \nonumber 
\end{eqnarray}

\noindent
The electromagnetic field components for a LNRO are (see Appendix C)

\begin{eqnarray}
\label{HT5}
E_r &=& q_e u^2 + 4 (J \mu + q q_e) P_2 u^5 , \nonumber \\
E_{\theta} &=& 3 (2 J \mu + q q_e) u^5 \cos{\theta} \sin{\theta} , \\
H_r &=& (2 \mu u^3 + 3  (\mu M - J q_e) u^4 + 5 \mu q P_2 u^6) \cos{\theta} , \nonumber \\
H_{\theta} &=& (\mu u^3 + (2 \mu M - 3 J q_e) u^4 + \mu q (3 P_2 + 1) u^6) \sin{\theta} . \nonumber
\end{eqnarray}

\noindent
The metric (\ref{HT}) is solution of the EME up to the third order in following parameters $ (J, \, \mu , q) $. 
The first terms in (\ref{HT5}) for $ E_{r} $, for $ H_{r} $ and $ H_{\theta} $ are the electric field, and magnetic dipole common formulas, respectively.

\section{Comparison Among the Spacetimes}

In this section we compare the spacetimes obtained in the previous sections. The principal characteristics of these metrics are

\begin{itemize}
\item 
These spacetimes posses four parameters, namely, mass, angular momentum, mass quadrupole, electric charge and magnetic dipole.
\item 
These metrics are asymptotically flat.
\item
The Lense-Thirring metric, which can be deduced from the Kerr metric, in a series expansion to the second order of $M$ and to the first order of $J$ is contained in all spacetimes.
\item
The Reissner-Nordstr\"om metric in a series expansion to the second order of $M$ and $q_e$ is contained in all spacetimes.
\item 
The four potentials of these metrics contain the first order of the charge and magnetic dipole of the objects.
\item 
The electric field corresponds to the first order expected for a charged object, in the same way that the magnetic field corresponds to the first order expected 
for an object with magnetic dipole.
\end{itemize}

\noindent
Let us see the Taylor series expressions of the metrics up to the fourth order in $ M $, second order in $ q_e $ and $ \mu $, first order in $ J, \, S, \, M_2 $ and $ q $.

\noindent
From (\ref{Papapetrou9}), the Hoenselaers-Perj\'es metric with charge and magnetic dipole is

\begin{eqnarray}
\label{SM1}
V &=& 1 - 2 M u + qe^2 u^2 - 2 M_2 p2 u^3 + \mu^2 u^4 \cos^2{\theta} , \nonumber \\
W &=& - 2 S u \sin^2{\theta} , \nonumber \\
X &=& 1 + 2 M u + (4 M^2 - q_e^2) u^2 + 2 (4 M^3 + M_2 P_2) u^3 \\
&+& \frac{1}{3} (48 M^4 + \mu^2 (- 6 P_2^2 + 2 P_2 + 1)) u^4 , \nonumber \\
Y &=& r^2 \left(1 + 2 M_2 P_2 u^3 + \frac{1}{3} \mu^2 (- 6 P_2^2 + 2 P_2 + 1) u^4 \right) , \nonumber \\
Z &=& r^2 \left(1 + 2 M_2 P_2 u^3 - \frac{1}{3} \mu^2 (2 P_2 + 1) u^4 \right) \sin^2{\theta} . \nonumber
\end{eqnarray}

\noindent
From (\ref{Lewis10}), the Kerr-Newman with magnetic dipole and mass quadrupole is

\begin{eqnarray}
\label{SM2}
V &=& 1 - 2 M u + qe^2 u^2 - 2 q P_2 u^3 + \mu^2 u^4 \cos^2{\theta} , \nonumber \\
W &=& - 2 J u \sin^2{\theta} , \nonumber \\
X &=& 1 + 2 M u + (4 M^2 - q_e^2) u^2 + 2 (4 M^3 + q P_2) u^3 \\
&+& \frac{1}{4} (64 M^4 + \mu^2 (4 P_2 - 15)) u^4 , \nonumber \\
Y &=& r^2 \left(1 + 2 q P_2 u^3 + \frac{1}{3} \mu^2 (3 - P_2) u^4 \right) , \nonumber \\
Z &=& r^2 \left(1 + 2 q P_2 u^3 + \frac{1}{18} \mu^2 (19 - 7 P_2) u^4 \right) \sin {\theta}^2 . \nonumber 
\end{eqnarray}

\noindent
From (\ref{HT}), the Hartle-Thorne metric with charge and magnetic dipole is

\begin{eqnarray}
\label{SM3}
V &=& 1 - 2 M u + qe^2 u^2 + 2 q P_2 u^3 + \mu^2 u^4 \cos{\theta}  , \nonumber \\
W &=& - 2 J u \sin^2{\theta} , \nonumber \\
X &=& 1 + 2 M u + (4 M^2 - q_e^2) u^2 + 2 (4 M^3 - 2 q P_2) u^3 \\
&+& \frac{1}{3} (48 M^4 + 2 \mu^2 (P_2 - 1)) u^4, \nonumber \\
Y &=& r^2 \left(1 - 2 q P_2 u^3 - \frac{1}{3} \mu^2 P_2 u^4 \right) , \nonumber \\
Z &=& r^2 \left(1 - 2 q P_2 u - \frac{1}{3} \mu^2 P_2 u^2 \right) \sin^2{\theta} . \nonumber
\end{eqnarray}

\noindent
The first order of the four potential components for the metrics are

\begin{eqnarray}
\label{SM4}
A_t &=& - q_e u , \\
A_{\phi} &=& \mu u \sin^2{\theta} . \nonumber
\end{eqnarray}

\noindent
The components of the electric and magnetic fields at first order are 

\begin{eqnarray}
\label{SM5}
E_r &=& q_e u^2 , \nonumber \\
E_{\theta} &=& 0 , \\
H_{r} &=& 2 \mu u^3 \cos{\theta} , \nonumber \\
H_{\theta} &=& \mu u^3 \cos{\theta} . \nonumber
\end{eqnarray}

\noindent
From (\ref{SM1}, \ref{SM2}, \ref{SM3}, \ref{SM4}) it can be seen that the structure of the metrics is similar. The Hoenselaers-Perj\'es metric was deduced assuming the multipolar structure 
from the beginning. In \cite{Frutos2}, the multipole structure of a metric with five parameters, namely the mass, spin, mass quadrupole, spin octupole and mass hexadecapole was obtained using 
the formalism of Fodor et al. \cite{Fodor}. Moreover, the metrics derived in \cite{Frutos2} are contained in these new metrics without electric charge and magnetic dipole. 
Since the new metrics contain Reissner-Nordstr\"om, the remaining parameter $ \mu $ is the only one that remains to be interpreted, but from the structure of the electromagnetic four potential 
and the electromagnetic field it can be inferred that it corresponds to a magnetic dipole.

\section{Conclusions}

\noindent
In this contribution, we found three spacetimes, that can be employed in astrophysical calculations. The metrics include mass, rotation, quadrupole moment, charge, and magnetic dipole moment.
The first spacetime was found using the Hoenselaers-Perj\'es formalism. This metric is in canonical cylindrical coordinates and was transformed to spherical-like coordinates 
using the Kerr-Newman mapping. The second metric is a generalization of the Kerr-Newman metric. 
This is an excellent approximation, since the Kerr-Newman metric is exact, and the parameters $q$ and $\mu$ are small compared to the mass and spin for an ample range of real values. 
The last metric is an improvement of the Hartle-Thorne spacetime including charge, and magnetic dipole moment. 
It is important to explore the matching of interior with exterior solutions with magnetic dipole moment.
The metrics were found using REDUCE programs that are available on request.

\noindent
These metrics are relevant in astrophysics, since they include the approximation of the magnetic dipole moment, which makes them suitable for simulating real compact objects.
Our approximations can easily be determined with a higher order of precision by taking into account the higher order terms in the expansions. 
It can be done without changing significantly the form of the analytic spacetimes. Furthermore, they can easily be implemented numerically. 

\noindent
These metrics have potentialy many applications. 
For instance, they could be used to infer features of the structure of a compact object from astrophysical observations. 
The ISCO and the shadow of the compact object are another characteristics that can be calculated from these spacetimes. 
In addition, a ray tracing program including these spacetimes can be useful to study the chaotic behavior of geodesics, light scattering, determination of the shape of the neutron star, 
the thermal spectrum and pulse profiles.

\appendix

\section{Einstein-Maxwell Equations}

The EME are given by

\begin{eqnarray}
\label{EME}
R_{\mu \nu} - \frac{R}{2} g_{\mu \nu} &=& \kappa T_{\mu \nu} , \nonumber \\
\nabla_{\nu} [\sqrt{- g} F^{\mu \nu}] &=& 0 , \\
\partial_{\alpha} F_{\beta \gamma} + \partial_{\beta} F_{\gamma \alpha} + \partial_{\gamma} F_{\alpha \beta} &=& 0 , \nonumber
\end{eqnarray}

\noindent
where $ \kappa = 8 \pi G/c^4 $, $ g_{\mu \nu} $ are the components of the metric tensor, 
$ R_{\mu \nu} $ are the components of the Ricci tensor, $ R $ is the curvature scalar, and $ T_{\mu \nu} $ is the stress-energy tensor given by

$$ \frac{\kappa}{2} T_{\mu \nu} = F_{\mu \alpha} F_{\nu}^{\, \alpha} - \frac{1}{4} g_{\mu \nu} F_{\alpha \beta} F^{\alpha \beta} $$

\noindent
with $ F_{\mu \nu} = \partial_{\mu} A_{\nu} - \partial_{\nu} A_{\mu} $ with $ A_{\mu} = (- A_{t}, \, 0, \, 0, \, A_{\phi}) $.

\noindent
The EME for the LWP metric (\ref{Papapetrou}) are explicitly

\begin{eqnarray}
\label{EME1}
\rho^2 f \nabla^2 f &=& \rho^2 (\nabla f)^2 - f^4 (\nabla \omega)^2
+ 2 f [\rho^2 (\nabla A_{t})^2 + f^2 (\nabla A_{\phi} - \omega \nabla A_{t})^2] , \nonumber \\
\label{EME2}
0 &=& \nabla \cdot \left[\frac{f^2}{\rho^2} \nabla \omega - 4 \frac{f}{\rho^2} A_{t} (\nabla A_{\phi} - \omega \nabla A_{t}) \right] , \nonumber \\
\label{EMgrho}
\partial_{\rho} \gamma &=& \frac{\rho}{4 f^2} [(\partial_{\rho} f)^2 - (\partial_{z} f)^2]
- \frac{f^2}{4 \rho} [(\partial_{\rho} \omega)^2 - (\partial_{z} \omega)^2] 
\nonumber \\
&+& \frac{f}{\rho} [(\partial_{\rho} A_{\phi})^2 - (\partial_{z} A_{\phi})^2]
+ \left(\frac{f \omega^2}{\rho} - \frac{\rho}{f} \right) [(\partial_{\rho} A_{t})^2 - (\partial_{z} A_{t})^2] \nonumber \\
&-& \frac{2 f \omega}{\rho} ({\partial_{\rho} A_{t}} {\partial_{\rho} A_{\phi}} - {\partial_{z} A_{t}} {\partial_{z} A_{\phi}}) , \\
\label{EMgz}
\partial_{z} \gamma &=& \frac{\rho}{2 f^2} \partial_{\rho} f \partial_{z} f
- \frac{f^2}{2 \rho} \partial_{\rho} \omega \partial_{z} \omega
+ \frac{2 f}{\rho} \partial_{\rho} A_{\phi} \partial_{z} A_{\phi} \\
&+& 2 \left(\frac{f \omega^2}{\rho} - \frac{\rho}{f} \right) \partial_{\rho} A_{t} \partial_{z} A_{t}
- \frac{2 f \omega}{\rho} ({\partial_{\rho} A_{t}} {\partial_{z} A_{\phi}} + {\partial_{\rho} A_{\phi}} {\partial_{z} A_{t}}) \nonumber \\
0 &=& \nabla \cdot \left[\frac{f}{\rho^2} (\nabla A_{\phi} - \omega \nabla A_{t}) \right] , \nonumber \\
0 &=& \nabla \cdot \left[ \frac{1}{f} \nabla A_{t} - \frac{f \omega}{\rho^2} (\nabla A_{\phi} - \omega \nabla A_{t}) \right] . \nonumber 
\end{eqnarray}

\noindent
These expressions were verified using REDUCE programs that are available upon request.

\noindent
The EME for the Lewis metric (\ref{Lewis}) are solved introducing the Ansatz and solving for algebraic equations of the unknown variables. 
These EME are explicitly

\begin{eqnarray}
\label{EMLewis}
0 &=& Y ((X ({\cal R}{\partial_{r} Y} + V Y {\partial_{r} Z}) - {\cal R} Y {\partial_{r} X} - 2 W X Y {\partial_{r} W}) {\partial_{r} V} \\
&+& X (2 V (X ({\partial_{\theta} W})^2 + Y ({\partial_{r} W})^2) - Y Z ({\partial_{r} V})^2) + 2 {\cal R} X Y {\partial_{r} \partial_{r} V}) \nonumber \\
&-& X (X ({\cal R}{\partial_{\theta} Y} - V Y {\partial_{\theta} Z}) - {\cal R} Y {\partial_{\theta} X} + 2 W X Y {\partial_{\theta} W} + X Y Z {\partial_{\theta} V}) {\partial_{\theta} V} \nonumber \\
&+& 2 {\cal R} X^2 Y {\partial_{\theta} \partial_{\theta} V} - 4 ({\cal R} + W^2) X Y^2 ({\partial_{r} A_{t}})^2 - 4 ({\cal R} + W^2) X^2 Y ({\partial_{\theta} A_{t}})^2 \nonumber \\
&-& 4 V X Y^2 (V {\partial_{r} A_{\phi}} - 2 W {\partial_{r} A_{t}}) {\partial_{r} A_{\phi}} 
- 4 V X^2 Y (V {\partial_{\theta} A_{\phi}} - 2 W {\partial_{\theta} A_{t}}) {\partial_{\theta} A_{\phi}} , \nonumber \\
0 &=& Y ((X ({\cal R}{\partial_{r} Y} - V Y {\partial_{r} Z}) -  {\cal R} Y {\partial_{r} X}) {\partial_{r} W} + 2  {\cal R} X Y {\partial_{r} \partial_{r} W}) \\
&-& X (X ({\cal R}{\partial_{\theta} Y} + V Y {\partial_{\theta} Z}) - {\cal R} Y {\partial_{\theta} X}) {\partial_{\theta} W} + 2 {\cal R} X^2 Y {\partial_{\theta} \partial_{\theta} W} \nonumber \\
&-& X Y^2 (Z {\partial_{r} W} - 2 W {\partial_{r} Z}) {\partial_{r} V} -  X^2 Y (Z {\partial_{\theta} W} - 2 W {\partial_{\theta} Z}) {\partial_{\theta} V} \nonumber \\
&+& 4 W X Y^2 Z ({\partial_{r} A_{t}})^2 + 4 W X^2 Y Z ({\partial_{\theta} A_{t}})^2 \nonumber \\
&-& 4 X Y^2 (V W {\partial_{r} A_{\phi}} + 2 {\cal R}{\partial_{r} A_{t}} - 2 W^2 {\partial_{r} A_{t}}) {\partial_{r} A_{\phi}} \nonumber \\
&-& 4 X^2 Y (V W {\partial_{\theta} A_{\phi}} + 2 {\cal R}{\partial_{\theta} A_{t}} - 2 W^2 {\partial_{\theta} A_{t}}) {\partial_{\theta} A_{\phi}} , \nonumber \\
0 &=& X (V Y^2 (2 {\cal R}{\partial_{r} \partial_{r} Z} - V {\partial_{r} Z}^2 V) - {\cal R}^2 ({\partial_{r} Y})^2 + 2 {\cal R}^2 Y {\partial_{r} \partial_{r} Y}) \\
&-& {\cal R} Y ({\cal R}{\partial_{r} Y} + V Y {\partial_{r} Z}) {\partial_{r} X}
- {\cal R} (X ({\cal R}{\partial_{\theta} Y} - V Y {\partial_{\theta} Z}) + {\cal R} Y {\partial_{\theta} X}) {\partial_{\theta} X} \nonumber \\
&+& 2 {\cal R}^2 X Y {\partial_{\theta} \partial_{\theta} X} - 2 Y^2 (W ({\cal R}{\partial_{r} X} + 2 V X {\partial_{r} Z}) \nonumber \\
&-& ({\cal R} - 2 W^2) X {\partial_{r} W}) {\partial_{r} W} + 4 {\cal R} W X Y^2 {\partial_{r} \partial_{r} W} + 2 {\cal R} W X Y {\partial_{\theta} W} {\partial_{\theta} X} \nonumber \\
&-& Y^2 ({\cal R} Z {\partial_{r} X} - 2 W^2 X {\partial_{r} Z} + 4 W X Z {\partial_{r} W} + X Z^2 {\partial_{r} V}) {\partial_{r} V} \nonumber \\
&+& 2 {\cal R} X Y^2 Z {\partial_{r} \partial_{r} V} + {\cal R} X Y Z {\partial_{\theta} V} {\partial_{\theta} X} \nonumber \\
&-& 4 {\cal R} X Y^2 Z ({\partial_{r} A_{t}})^2 + 4 {\cal R} X^2 Y Z ({\partial_{\theta} A_{t}})^2 \nonumber \\
&+& 4 {\cal R} X Y^2 (V {\partial_{r} A_{\phi}} - 2 W {\partial_{r} A_{t}}) {\partial_{r} A_{\phi}} 
- 4 {\cal R} X^2 Y (V {\partial_{\theta} A_{\phi}} - 2 W {\partial_{\theta} A_{t}}) {\partial_{\theta} A_{\phi}} , \nonumber \\
0 &=& V (X (Y (2 {\cal R}{\partial_{r} \partial_{\theta} Z} - V {\partial_{\theta} Z} {\partial_{r} Z}) - {\cal R}{\partial_{r} Y} {\partial_{\theta} Z}) 
- {\cal R} Y {\partial_{\theta} X} {\partial_{r} Z}) \\
&-& 2 W Y ({\cal R}{\partial_{\theta} X} + V X {\partial_{\theta} Z}) {\partial_{r} W} - 2 X (W ({\cal R}{\partial_{r} Y} + V Y {\partial_{r} Z}) \nonumber \\
&-& ({\cal R} - 2 W^2) Y {\partial_{r} W}) {\partial_{\theta} W} + 4 {\cal R} W X Y {\partial_{r} \partial_{\theta} W} \nonumber \\
&-& Y ({\cal R} Z {\partial_{\theta} X} - W^2 X {\partial_{\theta} Z} + 2 W X Z {\partial_{\theta} W}) {\partial_{r} V} \nonumber \\
&-& X ({\cal R} Z {\partial_{r} Y} - W^2 Y {\partial_{r} Z} + 2 W Y Z {\partial_{r} W} + Y Z^2 {\partial_{r} V}) {\partial_{\theta} V} \nonumber \\
&+& 2 {\cal R} X Y Z {\partial_{r} \partial_{\theta} V} - 8 {\cal R} X Y Z {\partial_{\theta} A_{t}} {\partial_{r} A_{t}} 
- 8 {\cal R} W X Y {\partial_{r} A_{\phi}} {\partial_{\theta} A_{t}} \nonumber \\
&+& 8 {\cal R} X Y (V {\partial_{r} A_{\phi}} - W {\partial_{r} A_{t}}) {\partial_{\theta} A_{\phi}} , \nonumber \\
0 &=& X ({\cal R} ({\cal R}{\partial_{r} Y} - V Y {\partial_{r} Z}) {\partial_{r} Y} - V X Y (2 {\cal R}{\partial_{\theta} \partial_{\theta} Z} - V ({\partial_{\theta} Z})^2) \\
&-& 2 {\cal R}^2 Y {\partial_{r} \partial_{r} Y} + {\cal R} V X {\partial_{\theta} Y} {\partial_{\theta} Z}) + {\cal R}^2 Y {\partial_{r} X} {\partial_{r} Y} \nonumber \\
&+& {\cal R}^2  (Y {\partial_{\theta} X} + X {\partial_{\theta} Y}) {\partial_{\theta} X} 
-2 {\cal R}^2 X Y {\partial_{\theta} \partial_{\theta} X} - 2 {\cal R} W X Y {\partial_{r} W} {\partial_{r} Y} \nonumber \\
&+& 2 X^2 (W ({\cal R}{\partial_{\theta} Y} + 2 V Y {\partial_{\theta} Z}) - ({\cal R} - 2 W^2) Y {\partial_{\theta} W}) {\partial_{\theta} W} \nonumber \\
&-& 4 {\cal R} W X^2 Y {\partial_{\theta} \partial_{\theta} W} - {\cal R} X Y Z {\partial_{r} V} {\partial_{r} Y} \nonumber \\
&+& X^2 ({\cal R} Z {\partial_{\theta} Y} - 2 W^2 Y {\partial_{\theta} Z} + 4 W Y Z {\partial_{\theta} W} + Y Z^2 {\partial_{\theta} V}) {\partial_{\theta} V} \nonumber \\
&-& 2 {\cal R} X^2 Y Z {\partial_{\theta} \partial_{\theta} V} - 4 {\cal R} X Y^2 Z ({\partial_{r} A_{t}})^2 + 4 {\cal R} X^2 Y Z ({\partial_{\theta} A_{t}})^2 \nonumber \\
&+& 4 {\cal R} X Y^2 (V {\partial_{r} A_{\phi}} - 2 W {\partial_{r} A_{t}}) {\partial_{r} A_{\phi}} 
- 4 {\cal R} X^2 Y (V {\partial_{\theta} A_{\phi}} - 2 W {\partial_{\theta} A_{t}}) {\partial_{\theta} A_{\phi}} , \nonumber \\
0 &=& X (Y (Y (2 {\cal R}{\partial_{r} \partial_{r} Z} - V {\partial_{r} Z}^2) - V X {\partial_{\theta} Z}^2 + 2 {\cal R} X {\partial_{\theta} \partial_{\theta} Z} 
+ {\cal R}{\partial_{r} Y} {\partial_{r} Z}) \nonumber \\
&-& {\cal R} X {\partial_{\theta} Y} {\partial_{\theta} Z}) 
- {\cal R} Y^2 {\partial_{r} X} {\partial_{r} Z} + {\cal R} X Y {\partial_{\theta} X} {\partial_{\theta} Z} \\
&+& 2 X Y^2 (Z {\partial_{r} W} - W {\partial_{r} Z}) {\partial_{r} W} 
+ 2 X^2 Y (Z {\partial_{\theta} W} - W {\partial_{\theta} Z}) {\partial_{\theta} W} \nonumber \\
&+& X Y^2 Z {\partial_{r} V} {\partial_{r} Z} + X^2 Y Z {\partial_{\theta} V} {\partial_{\theta} Z} \nonumber \\ 
&+& 4 X Y^2 Z^2 ({\partial_{r} A_{t}})^2 + 4 X^2 Y Z^2 ({\partial_{\theta} A_{t}})^2 \nonumber \\  
&+& 4 X Y^2 (({\cal R} + W^2) {\partial_{r} A_{\phi}} + 2 W Z {\partial_{r} A_{t}}) {\partial_{r} A_{\phi}} \nonumber \\ 
&+& 4 X^2 Y (({\cal R} + W^2) {\partial_{\theta} A_{\phi}} + 2 W Z {\partial_{\theta} A_{t}}) {\partial_{\theta} A_{\phi}} , \nonumber \\
0 &=& Y ((X (({\cal R} + W^2) Y {\partial_{r} Z} + {\cal R} Z) {\partial_{r} Y} - {\cal R} Y Z {\partial_{r} X} - 2 W X Y Z {\partial_{r} W} \\
&-& X Y Z^2 {\partial_{r} V}) {\partial_{r} A_{t}} + 2 {\cal R} X Y Z {\partial_{r} \partial_{r} A_{t}}) \nonumber \\
&+& X (X (({\cal R} + W^2) Y {\partial_{\theta} Z} - {\cal R} Z {\partial_{\theta} Y}) + {\cal R} Y Z {\partial_{\theta} X} - 2 W X Y Z {\partial_{\theta} W} \nonumber \\
&-& X Y Z^2 {\partial_{\theta} V}) {\partial_{\theta} A_{t}} + 2 {\cal R} X^2 Y Z {\partial_{\theta} \partial_{\theta} A_{t}} \nonumber \\
&+& Y (W (X ({\cal R} {\partial_{r} Y} - V Y {\partial_{r} Z}) - {\partial_{r} X} {\cal R} Y) + 2 ({\cal R} - W^2) X Y {\partial_{r} W} \nonumber \\
&-& W X Y Z {\partial_{r} V}) {\partial_{r} A_{\phi}} + 2 {\cal R} W X Y^2 {\partial_{r} \partial_{r} A_{\phi}} \nonumber \\
&-& X (W (X ({\cal R}{\partial_{\theta} Y} + V Y {\partial_{\theta} Z}) - {\cal R} Y {\partial_{\theta} X}) - 2 ({\cal R} - W^2) X Y {\partial_{\theta} W} \nonumber \\
&+& W X Y Z {\partial_{\theta} V}) {\partial_{\theta} A_{\phi}} + 2 {\cal R} W X^2 Y {\partial_{\theta} \partial_{\theta} A_{\phi}} , \nonumber \\
\label{EMLewis2}
0 &=& Y ((2 {\cal R} X Y {\partial_{r} W} - 2 W^2 X Y {\partial_{r} W} - {\cal R} W Y {\partial_{r} X} + {\cal R} W X {\partial_{r} Y} \\
&-& V W X Y {\partial_{r} Z} - W X Y Z {\partial_{r} V}) {\partial_{r} A_{t}} + 2 {\cal R} W X Y {\partial_{r} \partial_{r} A_{t}}) \nonumber \\
&+& X (2 {\cal R} X Y {\partial_{\theta} W} - 2 W^2 X Y {\partial_{\theta} W} + {\cal R} W Y {\partial_{\theta} X} - {\cal R} W X {\partial_{\theta} Y} \nonumber \\
&-& V W X Y {\partial_{\theta} Z} - W X Y Z {\partial_{\theta} V}) {\partial_{\theta} A_{t}} + 2 {\cal R} W X^2 Y {\partial_{\theta} \partial_{\theta} A_{t}} \nonumber \\
&-& Y (V (X ({\cal R}{\partial_{r} Y} - V Y {\partial_{r} Z}) - {\cal R} Y {\partial_{r} X} - 2 W X Y {\partial_{r} W}) \nonumber \\
&+& ({\cal R} + W^2) X Y {\partial_{r} V}) {\partial_{r} A_{\phi}} - 2 {\cal R} V X Y^2 {\partial_{r} \partial_{r} A_{\phi}} \nonumber \\
&+& X (V (X ({\cal R}{\partial_{\theta} Y} + V Y {\partial_{\theta} Z}) - {\cal R} Y {\partial_{\theta} X} + 2 W X Y {\partial_{\theta} W}) \nonumber \\
&-& ({\cal R} + W^2) X Y {\partial_{\theta} V}) {\partial_{\theta} A_{\phi}} - 2 {\cal R} V X^2 Y {\partial_{\theta} \partial_{\theta} A_{\phi}} , \nonumber 
\end{eqnarray}

\noindent
where $ {\cal R} = V Z + W^2 $. If one chooses $ A_{\mu} = (A_{t}, \, 0, \, 0, \, A_{\phi}) $, one has to change the signs of $ A_{t} \rightarrow - A_{t} $ in expressions (\ref{EME1}), 
and subsequent ones, as well as in (\ref{EMLewis}) and subsequent ones. These expressions were found using REDUCE programs that are available upon request.

\section{Ernst Equations}

\noindent
In 1968, Ernst reformulated the EME for the LWP metric (\ref{Papapetrou}) in a complex form using the potential \cite{Ernst1,Ernst2}

\begin{eqnarray}
\label{Ernstpot}
{\cal E} = (f - |\Phi|^2) + {\rm i} \varphi , 
\end{eqnarray}

\noindent
where the function $ {\cal E} $ is the Ernst potential, and 

\begin{eqnarray}
\label{Ernstpot2}
\Phi &=& A_t + {\rm i} {\tilde A}_{\phi} ,
\end{eqnarray}

\noindent
with $A_t$ and $A_{\phi}$ as the components of the electromagnetic four-potential. 
The function $ \varphi $ is the twist scalar, and $ {\tilde A}_{\phi} $ is an auxiliary potential, both functions are defined via

\begin{eqnarray}
\label{Ernstpot3}
{\hat{\bf{n}}} \times {\nabla A_{\phi}} &=& - \frac{\rho}{f} \nabla {\tilde A}_{\phi} + \omega {\hat{\bf{n}}} \times \nabla A_t,  \\
\label{Ernstpot4}
{\hat{\bf{n}}} \times {\nabla \omega} &=& - \frac{\rho}{f^2} [\nabla \varphi + 2 {\rm Im}(\Phi^{\star} \nabla \Phi)] , 
\end{eqnarray}

\noindent
where $ {\hat{\bf{n}}} $ is a unit vector in the azimuthal direction. From these equations, the EME can be rewritten as 

\noindent
\begin{eqnarray}
\label{Ernsteqs}
({\rm Re}{\cal E} + |\Phi|^2) \nabla^2 {\cal E} &=& (\nabla {\cal E} + 2 \Phi^{\star} \nabla \Phi) \cdot \nabla {\cal E} , \nonumber \\
({\rm Re}{\cal E} + |\Phi|^2) \nabla^2 {\Phi}   &=& (\nabla {\cal E} + 2 \Phi^{\star} \nabla \Phi) \cdot \nabla {\Phi} , \\
\partial_{\rho} \gamma &=& \frac{1}{4} \frac{\rho}{f^2} \left[ 
(\partial_{\rho}{\cal E} + 2 \Phi^{\star} \partial_{\rho}{\Phi})(\partial_{\rho}{\cal E}^{\star} + 2 \Phi \partial_{\rho}{\Phi^{\star}})
\right. \nonumber \\
&-& \left. (\partial_{z}{\cal E} + 2 \Phi^{\star} \partial_{z}{\Phi})(\partial_{z}{\cal E}^{\star} + 2 \Phi \partial_{z}{\Phi^{\star}})
\right] \nonumber \\
&-& \frac{\rho}{f} (\partial_{\rho}{\Phi} \partial_{\rho}{\Phi^{\star}} - \partial_{z}{\Phi} \partial_{z}{\Phi^{\star}}) , 
\nonumber \\
\partial_{z} \gamma &=& \frac{1}{4} \frac{\rho}{f^2} \left[ 
(\partial_{\rho}{\cal E} + 2 \Phi^{\star} \partial_{\rho}{\Phi})(\partial_{z}{\cal E}^{\star} + 2 \Phi \partial_{z}{\Phi^{\star}}) 
\right. \nonumber \\
&-& \left. (\partial_{z}{\cal E} + 2 \Phi^{\star} \partial_{z}{\Phi})(\partial_{\rho}{\cal E}^{\star} + 2 \Phi \partial_{\rho}{\Phi^{\star}})
\right] \nonumber \\
&-& \frac{\rho}{f} (\partial_{\rho}{\Phi} \partial_{z}{\Phi^{\star}} - \partial_{z}{\Phi} \partial_{\rho}{\Phi^{\star}}) . \nonumber
\end{eqnarray}

\noindent
These equations are the Ernst equations. Ernst introduced new functions $q , \, \xi$ such that

\begin{eqnarray}
\label{Ernstfuncs}
{\cal E} &=& \frac{\xi - 1}{\xi + 1} , \\  
\Phi &=& \frac{q}{\xi + 1} \nonumber
\end{eqnarray}

\noindent
Under the transformation (\ref{Ernstfuncs}), the EME become

\begin{eqnarray}
\label{Ernstfuncs2}
(\xi \xi^{\star} + q q^{\star} - 1) \nabla^2 \xi &=& 2[\xi \nabla \xi + q \nabla q] \cdot \nabla \xi \\
(\xi \xi^{\star} + q q^{\star} - 1) \nabla^2 q   &=& 2[\xi \nabla \xi + q \nabla q] \cdot \nabla q , \nonumber \\
\partial_{\rho} \gamma &=& \frac{\rho}{|\xi + 1|^6 f^2} ((q^{\star} ((\xi + 1) \partial_{\rho}{q} - q \partial_{\rho}{\xi}) 
+ (\xi^{\star} + 1) \partial_{\rho}{\xi}) \nonumber \\
&\times& (q ((\xi^{\star} + 1) \partial_{\rho}{q^{\star}} - q^{\star} \partial_{\rho}{\xi^{\star}}) + (\xi + 1) \partial_{\rho}{\xi^{\star}}) \nonumber \\
&-& (q^{\star} ((\xi + 1) \partial_{z}{q} - q \partial_{z}{\xi}) + (\xi^{\star} + 1) \partial_{z}{\xi}) \nonumber \\
&\times& (q ((\xi^{\star} + 1) \partial_{z}{q^{\star}} - q^{\star} \partial_{z}{\xi^{\star}}) + (\xi + 1) \partial_{z}{\xi^{\star}}) \nonumber \\
&-& f |\xi + 1|^2 (((\xi + 1) \partial_{\rho}{q} - q \partial_{\rho}{\xi}) ((\xi^{\star} + 1) \partial_{\rho}{q^{\star}} - q^{\star} \partial_{\rho}{\xi^{\star}}) \nonumber \\
&+& ((\xi + 1) \partial_{z}{q} - q \partial_{z}{\xi}) ((\xi^{\star} + 1) \partial_{z}{q^{\star}} - q^{\star} \partial_{z}{\xi^{\star}}))) , \nonumber \\
\partial_{z} \gamma &=& \frac{\rho}{|\xi + 1|^6 f^2} ((q^{\star} ((\xi + 1) \partial_{\rho}{q} - q \partial_{\rho}{\xi}) + (\xi^{\star} + 1) \partial_{\rho}{\xi}) \nonumber \\
&\times& (q ((\xi^{\star} + 1) \partial_{z}{q^{\star}} - q^{\star} \partial_{z}{\xi^{\star}}) + (\xi + 1) \partial_{z}{\xi^{\star}}) \nonumber \\
&-& (q^{\star} ((\xi + 1) \partial_{z}{q} - q \partial_{z}{\xi}) + (\xi^{\star} + 1) \partial_{z}{\xi}) \nonumber \\
&\times& (q ((\xi^{\star} + 1) \partial_{\rho}{q^{\star}} - q^{\star} \partial_{\rho}{\xi^{\star}}) + (\xi + 1) \partial_{\rho}{\xi^{\star}}) \nonumber \\
&-& f |\xi + 1|^2 (((\xi + 1) \partial_{\rho}{q} - q \partial_{\rho}{\xi}) ((\xi^{\star} + 1) \partial_{z}{q^{\star}} - q^{\star} \partial_{z}{\xi^{\star}}) \nonumber \\
&+& ((\xi + 1) \partial_{z}{q} - q \partial_{z}{\xi}) ((\xi^{\star} + 1) \partial_{\rho}{q^{\star}} - q^{\star} \partial_{\rho}{\xi^{\star}}))) . \nonumber
\end{eqnarray}

\noindent
The complex transformation (\ref{Ernstpot}) leads to powerful tools for solving the EME for the LWP spacetime under initial conditions 
\cite{BZ1,BZ2,Sibgatullin,HKX}.

\section{Determination of the Electromagnetic Field}

\noindent
To determine the electromagnetic field we use the Cartan formalism \cite{Harrison,Wald,EW,KPM}. For this purpose, local non-rotating observers are chosen. The metric LWP can be written as follows

\begin{eqnarray}
\label{PapapetrouMet}
{\rm d} s^2 &=& - F_2 {\rm d} t^2 + \frac{{\rm e}^{2 \gamma}}{f} ({\rm d} \rho^2 + {\rm d} z^2) + F_1 ({\rm d} \phi - \tilde{\omega} {\rm d} t)^2 \nonumber \\
&=& - (\omega^t)^2 + (\omega^{\rho})^2 + (\omega^{z})^2 + (\omega^{\phi})^2 , 
\end{eqnarray}

\noindent
where

\begin{eqnarray}
\label{definitions}
F_1 &=& \frac{\rho^2}{f} - f \omega^2 , \nonumber \\
\tilde{\omega} &=& - \frac{f \omega}{F_1} = - \frac{f^2 \omega}{\rho^2 - f^2 \omega^2} , \\
F_2 &=& \frac{\rho^2}{F_1} = \frac{\rho^2 f}{\rho^2 - f^2 \omega^2} . \nonumber
\end{eqnarray}
 
\noindent
The 1-forms for (\ref{PapapetrouMet}) are

\begin{eqnarray}
\label{forms1}
\omega^t &=& \sqrt{F_2} {\rm d} t , \nonumber \\
\omega^{\rho} &=& \frac{{\rm e}^{\gamma}}{\sqrt{f}} {\rm d} \rho , \nonumber \\
\omega^{z} &=& \frac{{\rm e}^{\gamma}}{\sqrt{f}} {\rm d} z , \\
\omega^{\phi} &=& \sqrt{F_1} ({\rm d} \phi - \tilde{\omega} {\rm d} t) . \nonumber
\end{eqnarray}

\noindent
The potential 1-form is

\begin{eqnarray}
\label{potential}
A = A_t {\rm d} t + A_{\phi} {\rm d} \phi , 
\end{eqnarray}

\noindent
where $ A_t = A_t(\rho, \, z) $ and $ A_{\phi} = A_{\phi}(\rho, \, z) $.

\noindent
The Faraday 2-form is

\begin{eqnarray}
\label{Faraday1}
F &=& {\rm d} A \\
&=& \partial_{\rho} A_t {\rm d} \rho \wedge {\rm d} t + \partial_{z} A_{\phi} {\rm d} z \wedge {\rm d} t 
+ \partial_{\rho} A_{\phi} {\rm d} \rho \wedge {\rm d} \phi + \partial_{z} A_{\phi} {\rm d} z \wedge {\rm d} \phi \nonumber \\
&=& (\partial_{\rho} A_t + \tilde{\omega} \partial_{\rho} A_{\phi}) {\rm d} \rho \wedge {\rm d} t 
+ (\partial_{z} A_t + \tilde{\omega} \partial_{z} A_{\phi}) {\rm d} z \wedge {\rm d} t \nonumber \\
&+& \partial_{\rho} A_{\phi} {\rm d} \rho \wedge ({\rm d} \phi - \tilde{\omega} {\rm d} t)
+ \partial_{z} A_{\phi} {\rm d} z \wedge ({\rm d} \phi - \tilde{\omega} {\rm d} t) \nonumber \\
&=& E_{\rho} \omega^{\rho} \wedge \omega^{t} + E_{z} \omega^z \wedge \omega^{t} - H_{z} \omega^{\rho} \wedge \omega^{\phi} + H_{\rho} \omega^{z} \wedge \omega^{\phi} , \nonumber
\end{eqnarray}

\noindent
where the orthonormal components of the electromagnetic field for a LNRO are

\begin{eqnarray}
\label{EM1}
E_{\rho} &=& {\sqrt{\frac{f}{F_2}}} {\rm e}^{- \gamma} (\partial_{\rho} A_t + \tilde{\omega} \partial_{\rho} A_{\phi}) 
= {\frac{\sqrt{\rho^2 - f^2 \omega^2}}{\rho}} {\rm e}^{- \gamma} \left(\partial_{\rho} A_t - \frac{f^2 \omega}{\rho^2 - f^2 \omega^2} \partial_{\rho} A_{\phi} \right) , \nonumber \\
E_{z} &=& {\sqrt{\frac{f}{F_2}}} {\rm e}^{- \gamma} (\partial_{z} A_t + \tilde{\omega} \partial_{z} A_{\phi}) 
= {\frac{\sqrt{\rho^2 - f^2 \omega^2}}{\rho}} {\rm e}^{- \gamma} \left(\partial_{z} A_t - \frac{f^2 \omega}{\rho^2 - f^2 \omega^2} \partial_{z} A_{\phi} \right) , \nonumber \\
H_{\rho} &=& {\sqrt{\frac{f}{F_1}}} {\rm e}^{- \gamma} \partial_{z} A_{\phi} = \frac{f {\rm e}^{- \gamma}}{\sqrt{\rho^2 - f^2 \omega^2}} \partial_{z} A_{\phi} , \\
H_{z} &=& - {\sqrt{\frac{f}{F_1}}} {\rm e}^{- \gamma} \partial_{\rho} A_{\phi} = - \frac{f {\rm e}^{- \gamma}}{\sqrt{\rho^2 - f^2 \omega^2}} \partial_{\rho} A_{\phi} . \nonumber
\end{eqnarray}

\noindent
In the case of the Lewis metric we have

\begin{eqnarray}
\label{LewisMet}
{\rm d} s^2 &=& - \left(\sqrt{V + \frac{W^2}{Z}} {\rm d} t \right)^2 + \left[\sqrt{X} {\rm d} r \right]^2 + \left[\sqrt{Y} {\rm d} \theta \right]^2 \nonumber \\
&+& \left(\sqrt{Z} {\rm d} \phi + \frac{W}{\sqrt{Z}} {\rm d} t \right)^2
\end{eqnarray}

\noindent
The 1-forms are

\begin{eqnarray}
\label{forms2}
\omega^{t} &=& \sqrt{V + \frac{W^2}{Z}} {\rm d} t , \nonumber \\
\omega^{r} &=& \sqrt{X} {\rm d} r , \\
\omega^{\theta} &=& \sqrt{Y} {\rm d} \theta , \nonumber \\
\omega^{\phi} &=& \sqrt{Z} \left( {\rm d} \phi + \frac{W}{Z} {\rm d} t \right) . \nonumber
\end{eqnarray}

\noindent
The potential 1-form is written as in (\ref{potential}), but where $ A_t = A_t(r, \, \theta) $ and $ A_{\phi} = A_{\phi}(r, \, \theta) $.

\noindent
The Faraday 2-form is

\begin{eqnarray}
\label{Faraday2}
F &=& {\rm d} A \\
&=& \partial_{r} A_t {\rm d} r \wedge {\rm d} t + \partial_{\theta} A_t {\rm d} \theta \wedge {\rm d} t + \partial_{r} A_{\phi} {\rm d} r \wedge {\rm d} \phi 
+ \partial_{\theta} A_{\phi} {\rm d} \theta \wedge {\rm d} \phi \nonumber \\
&=& \left(\partial_{r} A_t - \frac{W}{Z} \partial_{r} A_{\phi} \right) {\rm d} r \wedge {\rm d} t + \left(\partial_{\theta} A_t 
- \frac{W}{Z} \partial_{\theta} A_{\phi} \right) {\rm d} \theta \wedge {\rm d} t \nonumber \\
&+& \partial_{r} A_{\phi} {\rm d} r \wedge \left({\rm d} \phi + \frac{W}{Z} {\rm d} t \right) + \partial_{\theta} A_{\phi} {\rm d} \theta \wedge \left({\rm d} \phi 
+ \frac{W}{Z} {\rm d} t \right)\nonumber \\
&=& E_{r} \omega^{r} \wedge \omega^{t} + E_{\theta} \omega^{\theta} \wedge \omega^{t} - H_{\theta} \omega^{r} \wedge \omega^{\phi} + H_{r} \omega^{\theta} \wedge \omega^{\phi} , \nonumber
\end{eqnarray}

\noindent
where the orthonormal components of the electromagnetic field for a LNRO are

\begin{eqnarray}
\label{EM2}
E_{r} &=& \frac{1}{\sqrt{X (V Z + W^2)}} \left(\sqrt{Z} \partial_{r} A_t - \frac{W}{\sqrt{Z}} \partial_{r} A_{\phi} \right) , \nonumber \\
E_{\theta} &=& \frac{1}{\sqrt{Y (V Z + W^2)}} \left(\sqrt{Z} \partial_{\theta} A_t - \frac{W}{\sqrt{Z}} \partial_{\theta} A_{\phi} \right), \\
H_{r} &=& \frac{1}{\sqrt{Y Z}} \partial_{\theta} A_{\phi} , \nonumber \\
H_{\theta} &=& - \frac{1}{\sqrt{X Z}} \partial_{r} A_{\phi} . \nonumber
\end{eqnarray}

\noindent
For the Kerr-Newman metric the electromagnetic field for a LNRO are \cite{KPM,BSB}

\begin{eqnarray}
\label{EM3}
E_{K r} &=& \frac{q_e}{\rho^4 \sqrt{\Sigma}} (r^2 + a^2) (r^2 - a^2 \cos^2{\theta}) , \nonumber \\
E_{K \theta} &=& - 2 \frac{a^2 q_e r}{\rho^4} \sqrt{\frac{\Delta}{\Sigma}} \cos{\theta} \sin{\theta} , \\
H_{K r} &=& 2 \frac{a q_e r}{\rho^4 \sqrt{\Sigma}} (r^2 + a^2) \cos{\theta} , \nonumber \\
H_{K \theta} &=& \frac{a q_e}{\rho^4} \sqrt{\frac{\Delta}{\Sigma}} (r^2 - a^2 \cos^2{\theta}) \sin{\theta} ,  \nonumber
\end{eqnarray}

\noindent
where $ \Sigma = (r^2 + a^2)^2 - a^2 \Delta \sin^2{\theta} $.

\section{Hoenselaers-Perj\'es Relationships for the \\ Relativistic Multipoles}

The parameters $ P_i $ represent the mass or spin relativistic multipoles, if $i$ is even, it is a massive multipole, $i$ is odd, it is a spin multipole (complex value). 
The parameters $ Q_i $ represent the electromagnetic relativistic multipoles, if $i$ is even, the value is real, if $i$ is odd, it is a complex value. They are given by \cite{FCH,CGC}

\begin{eqnarray}
\label{RM}
P_0 &=& m_0 , \nonumber \\
P_1 &=& m_1 , \nonumber \\
P_2 &=& m_2 , \nonumber \\
P_3 &=& m_3 + \frac{1}{5} q^{\star}_0 S_{10} , \nonumber \\
P_4 &=& m_4 - \frac{1}{7} m_0^{\star} M_{20} + \frac{3}{35} q_1^{\star} S_{10} + \frac{1}{7} q_0^{\star} (3 S_{20} - 2 H_{20}) , \nonumber \\
P_5 &=& m_5 - \frac{1}{3} m_0^{\star} M_{30} - \frac{1}{21} m_1^{\star} M_{20} + \frac{1}{21} q_2^{\star} S_{10} + \frac{1}{21} q_1^{\star} (4 S_{20} - 3 H_{20}) \nonumber \\
&+& \frac{1}{21} q_0^{\star} ((q_0^{\star} q_0 - m_0^{\star} m_0) S_{10} + 14 S_{30} + 13 S_{21} - 7 H_{30}), \nonumber \\
P_6 &=& m_6 - \frac{5}{231} m_2^{\star} M_{20} - \frac{4}{33} m_1^{\star} M_{30} + \frac{1}{33} {m_0^{\star}}^2 m_0 M_{20} \nonumber \\
&-& \frac{1}{33} m_0^{\star} (18 M_{40} + 8M_{31}) + \frac{1}{33} q_3^{\star} S_{10} \nonumber \\
&+& \frac{1}{231} q_2^{\star} (25 S_{20} - 20 H_{20}) + \frac{2}{231} q_1^{\star} (35 S_{30} + 37 S_{21} - 21 H_{30}) \nonumber \\
&-& \frac{1}{1155} (37 q_1^{\star} m_0^{\star} + 13 q_0^{\star}  m_1^{\star}) m_0 S_{10} + \frac{1}{33} {q_0^{\star}}^2 (5 q_0 S_{20} - 4 m_0 Q_{20} + 3 q_1 S_{10}) \nonumber \\
&+& \frac{10}{231} q_1^{\star} q_0^{\star} q_0 S_{10} + \frac{2}{33} q_0^{\star} m_0^{\star} (2 m_0 H_{20} - 3 q_0 M_{20} - 2 m_1 S_{10}) \nonumber \\
&+& \frac{1}{33} q_0^{\star} (30 S_{40} + 32 S_{31} - 24 H_{31} - 12 H_{40}) \nonumber \\
Q_0 &=& q_0 , \\
Q_1 &=& q_1 , \nonumber \\
Q_2 &=& q_2 , \nonumber \\
Q_3 &=& q_3 - \frac{1}{5} m_0^{\star} H_{10} , \nonumber \\
Q_4 &=& q_4 + \frac{1}{7} q_0^{\star} Q_{20} - \frac{3}{35} m_1^{\star} H_{10} - \frac{1}{7} m_0^{\star} (3 H_{20} - 2 S_{20}) , \nonumber \\
Q_5 &=& q_5 + \frac{1}{3} q_0^{\star} Q_{30} + \frac{1}{21} q_1^{\star} Q_{20} - \frac{1}{21} m_2^{\star} H_{10} - \frac{1}{21} m_1^{\star} (4 H_{20} - 3 S_{20}) \nonumber \\
&+& \frac{1}{21} m_0^{\star} ((m_0^{\star} m_0 - q_0^{\star} q_0) H_{10} - 14 H_{30} - 13 H_{21} + 7 S_{30}) , \nonumber \\
Q_6 &=& q_6 + \frac{5}{231} q_2^{\star} Q_{20} + \frac{4}{33} q_1^{\star} Q_{30} + \frac{1}{33} {q_0^{\star}}^2 q_0 Q_{20} + \frac{1}{33} q_0^{\star} (18 Q_{40} + 8 Q_{31}) \nonumber \\
&-& \frac{1}{33} m_3^{\star} H_{10} - \frac{1}{231} m_2^{\star} (25 H_{20} - 20 S_{20}) \nonumber \\
&-& \frac{2}{231} m_1^{\star} (35 H_{30} + 37 H_{21} - 21 S_{30}) \nonumber \\
&-& \frac{1}{1155} (37 m_1^{\star} q_0^{\star} + 13 m_0^{\star} q_1^{\star}) q_0 H_{10} \nonumber \\
&+& \frac{1}{33} {m_0^{\star}}^2 (5 m_0 H_{20} - 4 q_0 M_{20} + 3 m_1 H_{10}) \nonumber \\
&+& \frac{10}{231} m_1^{\star} m_0^{\star} m_0 H_{10} 
+ \frac{2}{33} m_0^{\star} q_0^{\star} (2 q_0 S_{20} - 3 m_0 Q_{20} - 2 q_1 H_{10}) \nonumber \\
&-& \frac{1}{33} m_0^{\star} (30 H_{40} + 32 H_{31} - 24 S_{31} - 12 S_{40}) , \nonumber
\end{eqnarray}

\noindent
where $ m_i = a_{0 i} $, $ q_i = b_{0 i} $, and

\begin{eqnarray}
M_{ij} &=& m_{i} m_{j} - m_{i-1} m_{j+1} , \nonumber \\
Q_{ij} &=& q_{i} q_{j} - q_{i-1} q_{j+1} , \\
S_{ij} &=& m_{i} q_{j} - m_{i-1} q_{j+1} , \nonumber \\
H_{ij} &=& q_{i} m_{j} - q_{i-1} m_{j+1} . \nonumber
\end{eqnarray}

\noindent
Now, we choose the relativistic multipoles as follows 

\begin{eqnarray}
P_0 =& m , \qquad Q_0 &= q_e , \nonumber \\
P_1 =& {\rm i} S , \qquad Q_1 &= {\rm i} \mu, \\
P_2 =& M_2 , \qquad \! Q_2 &= 0 , \nonumber \\
P_n =& 0 , \; \qquad \; Q_n &= 0 , \; \forall \, n \ge 3 . \nonumber 
\end{eqnarray}

\noindent
From (\ref{RM}), one can invert the relationships to get the values $ m_{i} $ and $ q_{i} $

\begin{eqnarray}
\label{RM2}
m_0 &=& M , \nonumber \\
m_1 &=& {\rm i} S , \nonumber \\
m_2 &=& M_2 , \nonumber \\
m_3 &=& \frac{{\rm i}}{5} (M \mu q_{e} - q_{e}^2 S) , \nonumber \\
m_4 &=& \frac{1}{35} (5 M^2 M_2 + 3 M \mu^2 + 5 M S^2 - 15 M_2 q_{e}^2 - 8 \mu q_{e} S) , \nonumber \\
m_5 &=& \frac{{\rm i}}{21} (- 8 M M_2 S + 3 M_2 \mu q_e + \mu^2 S - S^3) , \nonumber \\ 
m_6 &=& \frac{1}{1155} (55 M^4 M_2 + 66 M^3 \mu^2 + 55 M^3 S^2 - 330 M^2 M_2 q_{e}^2 \nonumber \\
&-& 66 M^2 \mu q_{e} S - 255 M M_2^2 - 121 M \mu^2 q_{e}^2 - 110 M q_{e}^2 S^2 - 20 M_2 \mu^2 \nonumber \\
&+& 275 M_2 q_{e}^4 - 115 M_2 S^2 + 176 \mu q_{e}^3 S) , \nonumber \\ 
q_0 &=& q_e , \\
q_1 &=& {\rm i} \mu , \nonumber \\
q_2 &=& 0 , \nonumber \\
q_3 &=& \frac{{\rm i}}{5} M (M \mu - q_{e} S) , \nonumber \\
q_4 &=& \frac{1}{35} (- 10 M M_2 q_{e} + 8 M \mu S - 5 \mu^2 q_{e} - 3 q_{e} S^2) , \nonumber \\
q_5 &=& \frac{{\rm i}}{105} (9 M^4 \mu - 9 M^3 q_{e} S - 9 M^2 \mu q_{e}^2 - 25 M M_2 \mu + 9 M q_{e}^3 S \nonumber \\ 
&+& 10 M_2 q_{e} S + 5 \mu^3 - 5 \mu S^2) , \nonumber \\
q_6 &=& \frac{1}{1155} (- 220 M^3 M_2 q_{e} + 176 M^3 \mu S - 110 M^2 \mu^2 q_{e} - 121 M^2 q_{e} S^2 \nonumber \\ 
&+& 220 M M_2 q_{e}^3 - 66 M \mu q_{e}^2 S - 100 M_2^2 q_{e} - 135 M_2 \mu S + 55 \mu^2 q_{e}^3 \nonumber \\ 
&+& 66 q_{e}^3 S^2) . \nonumber
\end{eqnarray}

\noindent
Using the following recurrence relationships, we obtain the non-zero values of $ a_{ij} $ and $ b_{ij} $ for $ i + j \le 6 $ using (\ref{RM2})

\begin{eqnarray}
\label{RM3}
a_{20} &=& - \frac{1}{2} (a_{02} + a_{00}^2 a^{\star}_{00}) = - \frac{1}{2} (M^3 + M_2) , \nonumber \\
a_{21} &=& \frac{1}{2} (- 3 a_{03} - 4 a_{01} a_{00} a^{\star}_{00} - a_{00}^2 a^{\star}_{01}) = \frac{3 {\rm i}}{10} (- 5 M^2 S - M \mu q_{e} + q_{e}^2 S) , \nonumber \\
a_{22} &=& \frac{1}{2} (2 a_{20} a_{00} a^{\star}_{00} - 6 a_{04} - 5 a_{02} a_{00} a^{\star}_{00} - 4 a_{01}^2 a^{\star}_{00} - 4 a_{01} a_{00} a^{\star}_{01} - a_{00}^2 a^{\star}_{02}) \nonumber \\
&=& \frac{1}{70} (- 35 M^5 - 275 M^2 M_2 - 18 M \mu^2 - 30 M S^2 + 90 M_2 q_{e}^2 + 48 \mu q_{e} S) , \nonumber \\
a_{23} &=& \frac{1}{2} (2 a_{21} a_{00} a^{\star}_{00} + 2 a_{20} a_{01} a^{\star}_{00} + 2 a_{20} a_{00} a^{\star}_{01} - 10 a_{05} - 5 a_{03} a_{00} a^{\star}_{00} \nonumber \\
&-& 11 a_{02} a_{01} a^{\star}_{00} - 5 a_{02} a_{00} a^{\star}_{01} - 4 a_{01}^2 a^{\star}_{01} - 4 a_{01} a_{00} a^{\star}_{02} - a_{00}^2 a^{\star}_{03}) \nonumber \\
&=& \frac{{\rm i}}{210} (- 315 M^4 S - 237 M^3 \mu q_{e} + 237 M^2 q_{e}^2 S - 650 M M_2 S \nonumber \\
&+& 90 M \mu q_{e}^3 - 250 M_2 \mu q_{e} - 50 \mu^2 S - 90 q_{e}^4 S - 370 S^3) , \nonumber \\
a_{24} &=& \frac{1}{2} (2 a_{22} a_{00} a^{\star}_{00} + 2 a_{21} a_{01} a^{\star}_{00} + 2 a_{21} a_{00} a^{\star}_{01} + 2 a_{20} a_{02} a^{\star}_{00} + 2 a_{20} a_{01} a^{\star}_{01} \nonumber \\
&+& 2 a_{20} a_{00} a^{\star}_{02} - 15 a_{06} - 4 a_{04} a_{00} a^{\star}_{00} - 13 a_{03} a_{01} a^{\star}_{00} - 5 a_{03} a_{00} a^{\star}_{01} \nonumber \\
&-& 8 a_{02}^2 a^{\star}_{00} - 11 a_{02} a_{01} a^{\star}_{01} - 5 a_{02} a_{00} a^{\star}_{02} - 4 a_{01}^2 a^{\star}_{02} - 4 a_{01} a_{00} a^{\star}_{03} - a_{00}^2 a^{\star}_{04}) \nonumber \\
&=& \frac{1}{770} (- 385 M^7 - 4345 M^4 M_2 - 693 M^3 \mu^2 - 1265 M^3 S^2 \nonumber \\
&+& 3465 M^2 M_2 q_{e}^2 + 1606 M^2 \mu q_{e} S - 4500 M M_2^2 + 605 M \mu^2 q_{e}^2 \nonumber \\
&+& 242 M q_{e}^2 S^2 + 100 M_2 \mu^2 - 1375 M_2 q_{e}^4 - 2505 M_2 S^2 - 880 \mu q_{e}^3 S) , \nonumber \\
a_{40} &=& \frac{1}{8} (- a_{22} - 4 a_{20} a_{00} a^{\star}_{00} + a_{02} a_{00} a^{\star}_{00} - a_{01}^2 a^{\star}_{00} - a_{00}^2 a^{\star}_{20}) \nonumber \\
&=& \frac{1}{280} (105 M^5 + 260 M^2 M_2 + 9 M \mu^2 + 50 M S^2 - 45 M_2 q_{e}^2 - 24 \mu q_{e} S) , \nonumber \\
a_{41} &=& \frac{1}{8} (- 3 a_{23} - 6 a_{21} a_{00} a^{\star}_{00} - 6 a_{20} a_{01} a^{\star}_{00} - 4 a_{20} a_{00} a^{\star}_{01} + 3 a_{03} a_{00} a^{\star}_{00} \nonumber \\
&-& 3 a_{02} a_{01} a^{\star}_{00} + a_{02} a_{00} a^{\star}_{01} - a_{01}^2 a^{\star}_{01} - 4 a_{01} a_{00} a^{\star}_{20} - a_{00}^2 a^{\star}_{21}) \nonumber \\
&=& \frac{{\rm i}}{280}(525 M^4 S + 192 M^3 \mu q_{e} - 192 M^2 q_{e}^2 S + 290 M M_2 S - 45 M \mu q_{e}^3 \nonumber \\
&+& 125 M_2 \mu q_{e} + 25 \mu^2 S + 45 q_{e}^4 S + 150 S^3) , \nonumber \\
a_{42} &=& \frac{1}{8} (8 a_{40} a_{00} a^{\star}_{00} - 6 a_{24} - 7 a_{22} a_{00} a^{\star}_{00} - 10 a_{21} a_{01} a^{\star}_{00} - 6 a_{21} a_{00} a^{\star}_{01} \nonumber \\
&-& 2 a_{20}^2 a^{\star}_{00} - 7 a_{20} a_{02} a^{\star}_{00} - 6 a_{20} a_{01} a^{\star}_{01} + 2 a_{20} a_{00} a^{\star}_{20} - 4 a_{20} a_{00} a^{\star}_{02} \nonumber \\
&+& 6 a_{04} a_{00} a^{\star}_{00} - 3 a_{03} a_{01} a^{\star}_{00} + 3 a_{03} a_{00} a^{\star}_{01} - 3 a_{02}^2 a^{\star}_{00} - 3 a_{02} a_{01} a^{\star}_{01} \nonumber \\
&-& 5 a_{02} a_{00} a^{\star}_{20} + a_{02} a_{00} a^{\star}_{02} - 4 a_{01}^2 a^{\star}_{20} - a_{01}^2 a^{\star}_{02} - 4 a_{01} a_{00} a^{\star}_{21} - a_{00}^2 a^{\star}_{22}) \nonumber \\
&=& \frac{1}{3080} (3850 M^7 + 31405 M^4 M_2 + 3168 M^3 \mu^2 + 6380 M^3 S^2 \nonumber \\
&-& 15840 M^2 M_2 q_{e}^2 - 7260 M^2 \mu q_{e} S + 15810 M M_2^2 - 1815 M \mu^2 q_{e}^2 \nonumber \\
&-& 1188 M q_{e}^2 S^2 - 300 M_2 \mu^2 + 4125 M_2 q_{e}^4 + 7130 M_2 S^2 + 2640 \mu q_{e}^3 S) , \nonumber \\
a_{60} &=& \frac{1}{18} (- a_{42} - 2 a_{40} a_{00} a^{\star}_{00} + a_{22} a_{00} a^{\star}_{00} - 2 a_{21} a_{01} a^{\star}_{00} - 7 a_{20}^2 a^{\star}_{00} \nonumber \\
&+& a_{20} a_{02} a^{\star}_{00} - 4 a_{20} a_{00} a^{\star}_{20} + a_{02} a_{00} a^{\star}_{20} - a_{01}^2 a^{\star}_{20} - a_{00}^2 a^{\star}_{40}) \nonumber \\
&=& \frac{1}{18480} (- 5775 M^7 - 24035 M^4 M_2 - 1419 M^3 \mu^2 - 6710 M^3 S^2 \nonumber \\
&+& 7095 M^2 M_2 q_{e}^2 + 2772 M^2 \mu q_{e} S - 9120 M M_2^2 + 605 M \mu^2 q_{e}^2 \nonumber \\
&+& 1012 M q_{e}^2 S^2 + 100 M_2 \mu^2 - 1375 M_2 q_{e}^4 - 2890 M_2 S^2 - 880 \mu q_{e}^3 S) , \nonumber \\
b_{20} &=& - \frac{1}{2} (b_{02} + b_{00}^2 b^{\star}_{00}) = - \frac{1}{2} q_{e}^3 , \\
b_{21} &=& \frac{1}{2} (- 3 b_{03} - 4 b_{01} b_{00} b^{\star}_{00} - b_{00}^2 b^{\star}_{01}) = \frac{3 {\rm i}}{10} (- M^2 \mu + M q_{e} S - 5 \mu q_{e}^2) , \nonumber \\
b_{22} &=& \frac{1}{2} (2 b_{20} b_{00} b^{\star}_{00} - 6 b_{04} - 5 b_{02} b_{00} b^{\star}_{00} - 4 b_{01}^2 b^{\star}_{00} - 4 b_{01} b_{00} b^{\star}_{01} - b_{00}^2 b^{\star}_{02}) \nonumber \\
&=& \frac{1}{70} (60 M M_2 q_{e} - 48 M \mu S + 30 \mu^2 q_{e} - 35 q_{e}^5 + 18 q_{e} S^2) , \nonumber \\
b_{23} &=& \frac{1}{2} (2 b_{21} b_{00} b^{\star}_{00} + 2 b_{20} b_{01} b^{\star}_{00} + 2 b_{20} b_{00} b^{\star}_{01} - 10 b_{05} - 5 b_{03} b_{00} b^{\star}_{00} \nonumber \\
&-& 11 b_{02} b_{01} b^{\star}_{00} - 5 b_{02} b_{00} b^{\star}_{01} - 4 b_{01}^2 b^{\star}_{01} - 4 b_{01} b_{00} b^{\star}_{02} - b_{00}^2 b^{\star}_{03}) \nonumber \\
&=& \frac{{\rm i}}{210} (- 90 M^4 \mu + 90 M^3 q_{e} S - 57 M^2 \mu q_{e}^2 + 250 M M_2 \mu + 57 M q_{e}^3 S \nonumber \\
&-& 100 M_2 q_{e} S - 470 \mu^3 - 315 \mu q_{e}^4 + 50 \mu S^2) , \nonumber \\
b_{24} &=& \frac{1}{2} (2 b_{22} b_{00} b^{\star}_{00} + 2 b_{21} b_{01} b^{\star}_{00} + 2 b_{21} b_{00} b^{\star}_{01} + 2 b_{20} b_{02} b^{\star}_{00} + 2 b_{20} b_{01} b^{\star}_{01} \nonumber \\
&+& 2 b_{20} b_{00} b^{\star}_{02} - 15 b_{06} - 4 b_{04} b_{00} b^{\star}_{00} - 13 b_{03} b_{01} b^{\star}_{00} - 5 b_{03} b_{00} b^{\star}_{01} \nonumber \\
&-& 8 b_{02}^2 b^{\star}_{00} - 11 b_{02} b_{01} b^{\star}_{01} - 5 b_{02} b_{00} b^{\star}_{02} - 4 b_{01}^2 b^{\star}_{02} - 4 b_{01} b_{00} b^{\star}_{03} - b_{00}^2 b^{\star}_{04}) \nonumber \\
&=& \frac{1}{770} (1100 M^3 M_2 q_{e} - 880 M^3 \mu S + 858 M^2 \mu^2 q_{e} + 605 M^2 q_{e} S^2 \nonumber \\
&+& 110 M M_2 q_{e}^3 - 946 M \mu q_{e}^2 S + 500 M_2^2 q_{e} + 675 M_2 \mu S - 55 \mu^2 q_{e}^3 \nonumber \\
&-& 385 q_{e}^7 + 33 q_{e}^3 S^2) , \nonumber \\
b_{40} &=& \frac{1}{8} (- b_{22} - 4 b_{20} b_{00} b^{\star}_{00} + b_{02} b_{00} b^{\star}_{00} - b_{01}^2 b^{\star}_{00} - b_{00}^2 b^{\star}_{20}) \nonumber \\
&=& \frac{1}{280} (- 30 M M_2 q_{e} + 24 M \mu S + 20 \mu^2 q_{e} + 105 q_{e}^5 - 9 q_{e} S^2) , \nonumber \\
b_{41} &=& \frac{1}{8} (- 3 b_{23} - 6 b_{21} b_{00} b^{\star}_{00} - 6 b_{20} b_{01} b^{\star}_{00} - 4 b_{20} b_{00} b^{\star}_{01} + 3 b_{03} b_{00} b^{\star}_{00} \nonumber \\
&-& 3 b_{02} b_{01} b^{\star}_{00} + b_{02} b_{00} b^{\star}_{01} - b_{01}^2 b^{\star}_{01} - 4 b_{01} b_{00} b^{\star}_{20} - b_{00}^2 b^{\star}_{21}) \nonumber \\
&=& \frac{{\rm i}}{280} (45 M^4 \mu - 45 M^3 q_{e} S + 102 M^2 \mu q_{e}^2 - 125 M M_2 \mu - 102 M q_{e}^3 S \nonumber \\
&+& 50 M_2 q_{e} S + 200 \mu^3 + 525 \mu q_{e}^4 - 25 \mu S^2) , \nonumber \\
b_{42} &=& \frac{1}{8} (8 b_{40} b_{00} b^{\star}_{00} - 6 b_{24} - 7 b_{22} b_{00} b^{\star}_{00} - 10 b_{21} b_{01} b^{\star}_{00} - 6 b_{21} b_{00} b^{\star}_{01} \nonumber \\
&-& 2 b_{20}^2 b^{\star}_{00} - 7 b_{20} b_{02} b^{\star}_{00} - 6 b_{20} b_{01} b^{\star}_{01} + 2 b_{20} b_{00} b^{\star}_{20} - 4 b_{20} b_{00} b^{\star}_{02} \nonumber \\
&+& 6 b_{04} b_{00} b^{\star}_{00} - 3 b_{03} b_{01} b^{\star}_{00} + 3 b_{03} b_{00} b^{\star}_{01} - 3 b_{02}^2 b^{\star}_{00} - 3 b_{02} b_{01} b^{\star}_{01} \nonumber \\
&-& 5 b_{02} b_{00} b^{\star}_{20} + b_{02} b_{00} b^{\star}_{02} - 4 b_{01}^2 b^{\star}_{20} - b_{01}^2 b^{\star}_{02} - 4 b_{01} b_{00} b^{\star}_{21} - b_{00}^2 b^{\star}_{22}) \nonumber \\
&=& \frac{1}{3080} (- 3300 M^3 M_2 q_{e} + 2640 M^3 \mu S - 2112 M^2 \mu^2 q_{e} - 1815 M^2 q_{e} S^2 \nonumber \\
&-& 3960 M M_2 q_{e}^3 + 5280 M \mu q_{e}^2 S - 1500 M_2^2 q_{e} - 2025 M_2 \mu S - 880 \mu^2 q_{e}^3 \nonumber \\
&+& 3850 q_{e}^7 - 1188 q_{e}^3 S^2) , \nonumber \\
b_{60} &=& \frac{1}{18} (- b_{42} - 2 b_{40} b_{00} b^{\star}_{00} + b_{22} b_{00} b^{\star}_{00} - 2 b_{21} b_{01} b^{\star}_{00} - 7 b_{20}^2 b^{\star}_{00} \nonumber \\
&+& b_{20} b_{02} b^{\star}_{00} - 4 b_{20} b_{00} b^{\star}_{20} + b_{02} b_{00} b^{\star}_{20} - b_{01}^2 b^{\star}_{20} - b_{00}^2 b^{\star}_{40}) \nonumber \\
&=& \frac{1}{18480} (1100 M^3 M_2 q_{e} - 880 M^3 \mu S + 88 M^2 \mu^2 q_{e} + 605 M^2 q_{e} S^2 \nonumber \\
&+& 2530 M M_2 q_{e}^3 - 2112 M \mu q_{e}^2 S + 500 M_2^2 q_{e} + 675 M_2 \mu S - 3080 \mu^2 q_{e}^3 \nonumber \\
&-& 5775 q_{e}^7 + 759 q_{e}^3 S^2) . \nonumber
\end{eqnarray}

\noindent
As already mentioned, not all terms in (\ref{RM2}) and (\ref{RM3}) held to satisfy the EME.

\section {Acknowledgement}

\noindent
Many thanks to Etevaldo dos Santos Costa Filho for the fruitful discussion that led to an improvement of the article.

\end{document}